\documentclass[a4paper,11pt]{article}

\usepackage{jheppub}
\usepackage{lineno}
\usepackage{physics}
\usepackage{xcolor}
\usepackage{CJK}
\usepackage{comment}
\usepackage[normalem]{ulem}

\newcommand{\jp}[1]{%
  {\begin{CJK}{UTF8}{min}#1\end{CJK}}%
}

\newcommand{\fix}[1]{\textcolor{red}{\jp{[#1]}}}

\preprint{%
\begin{flushright}
UT-Komaba/26-5\\
RIKEN-iTHEMS-Report-26
\end{flushright}
}

\title{$\mathrm{SL}(2,\mathbb{Z})$ Theta Subgroup Structure of Maxwell theory in the Lattice Villain Hamiltonian Formulation}

\author[a,b]{Shoto Aoki}
\emailAdd{shoto.aoki@riken.jp}
\affiliation[a]{Interdisciplinary Theoretical and Mathematical Sciences Program (iTHEMS), RIKEN, Wako 351-0198, Japan}

\affiliation[b]{Department of Physics, University of California, Berkeley, CA 94720, USA}

\author[c]{Yoshio Kikukawa}
\emailAdd{kikukawa@hep1.c.u-tokyo.ac.jp}
\affiliation[c]{Graduate School of Arts and Sciences, University of Tokyo, Komaba, Meguro-ku, Tokyo 153-8902, Japan}

\author[c]{Toshinari Takemoto}
\emailAdd{takemoto@hep1.c.u-tokyo.ac.jp}

\abstract{
We study the duality structure of lattice Maxwell theory with a theta term in the Hamiltonian Villain formulation. Reflecting the fact that odd-level Chern--Simons theory depends on a choice of
spin structure, a complete realization of the full $\mathrm{SL}(2,\mathbb{Z})$
structure would require fermionic degrees of freedom.
 We therefore restrict our analysis to the bosonic theory and focus on the theta subgroup generated by the $\mathcal{S}$ and $\mathcal{T}^{2}$ transformations. We construct these transformations at the operator level and show that they realize the theta subgroup structure of $\mathrm{SL}(2,\mathbb Z)$. We also extend the analysis to sectors with electric and magnetic charges, introduced as violations of the Gauss-law constraint and the Bianchi identity, respectively. We show that the $\mathcal S$ transformation exchanges electric and magnetic charges, while the $\mathcal T^2$ transformation realizes the Witten effect. Finally, we discuss a related non-invertible defect obtained by gauging a $\mathbb Z_N$ subgroup of the global $\mathrm{U}(1)$ $1$-form symmetry, and show that its fusion rule reproduces the expected Tambara--Yamagami structure.
}

\begin{document}
\maketitle
\flushbottom

\section{Introduction}
\label{sec:intro}
Maxwell theory is one of the most fundamental quantum field theories and provides a basic framework in which electric-magnetic duality can be studied in a precise manner. 
At the classical level, Maxwell's equations exhibit a striking symmetry between electric and magnetic fields. 
In the quantum theory, this symmetry is promoted to a nontrivial duality structure relating descriptions with different coupling constants. 
When the inverse gauge coupling $\beta$ and the topological angle $\theta$ are
combined into the complex coupling
\begin{align}
\tau = \frac{\theta}{2\pi} + 2\pi i \beta ,
\end{align}
the theory is expected to have a duality that identifies theories with
different values of $\tau$. The $\mathrm{SL}(2,\mathbb{Z})$ duality is
generated by the $\mathcal{S}$ and $\mathcal{T}$ transformations, which act on
the coupling as
\begin{align}
\label{eq:coupling transformation}
\mathcal{S}: \tau &\longrightarrow -\frac{1}{\tau},\\
\mathcal{T}: \tau &\longrightarrow \tau + 1 .
\end{align}
The $\mathcal{S}$ transformation exchanges electric and magnetic variables and relates the weakly coupled and strongly coupled regimes, while the $\mathcal{T}$ transformation shifts the $\theta$ angle by $2\pi$. 

The significance of the $\mathrm{SL}(2,\mathbb{Z})$ structure is not limited
to Maxwell theory.  Electric-magnetic duality has been a central theme in
four-dimensional gauge theory, from the structure of magnetic charges and the
Montonen--Olive conjecture in non-Abelian gauge theories
\cite{Goddard:1976qe,Montonen:1977sn} to supersymmetric realizations in
$\mathcal N=4$ super Yang--Mills theory
\cite{Witten:1978mh,Osborn:1979tq} and Seiberg--Witten theory
\cite{Seiberg:1994aj}. 
At the same time, Witten showed that in Abelian gauge theory on general
four-dimensional manifolds, the $\mathrm{SL}(2,\mathbb{Z})$ structure acts
nontrivially on the topological sectors, with the partition function
transforming as a modular form rather than as an invariant function
\cite{Witten:1995gf}.
These examples show that the $\mathrm{SL}(2,\mathbb{Z})$ structure provides
an important viewpoint on global and non-perturbative aspects of gauge theories,
including charge lattices, topological sectors, and strong-weak coupling
dualities.

Attempts have been made to realize these dualities in lattice models based on the modified Villain formulation \cite{Anosova:2022cjm,Anosova:2022yqx,Honda:2020txe,Hayashi:2022fkw,Katayama:2025pmz,Aoki:2026pvq}. In the Villain formulation \cite{Villain:1974ir} of $\mathrm{U}(1)$ gauge theory, an $\mathbb{R}$-valued gauge field is assigned to links, together with an auxiliary $\mathbb{Z}$-valued $2$-form field defined on plaquettes. The $\mathbb{Z}$-valued field encodes the compactness of the $\mathrm{U}(1)$ gauge field through a $\mathbb{Z}$ $1$-form gauge redundancy, while simultaneously making topological sectors manifest. 
This formulation also allows monopoles to be introduced naturally, in contrast
to conventional lattice formulations, where their description requires singular configurations \cite{DeGrand:1980eq}. In the modified Villain formulation, one further imposes a no-monopole condition so that the magnetic $2$-form field retains only topological flux information.

Recently, it has been shown that Euclidean lattice Maxwell theory with an ultra-local action in the modified Villain formulation possesses an exact $\mathrm{SL}(2,\mathbb{Z})$ duality \cite{Aoki:2026pvq}. The $\mathcal{S}$ transformation is essentially realized through the Poisson summation formula, while the $\mathcal{T}$ transformation corresponds to shifting the $\theta$ angle by $2\pi$. Wilson and 't Hooft loop operators also transform nontrivially under these duality transformations, acquiring phase factors determined by their linking structure. Remarkably, the resulting $\mathrm{SL}(2,\mathbb{Z})$ structure implies statistics transmutation.

Considerable progress has also been made in the understanding of the Hamiltonian formulation based on the Villain formalism \cite{Fazza:2022fss,Yoneda:2022qpj,Jacobson:2024hov}. In particular, the authors of \cite{Fazza:2022fss} constructed a Hamiltonian Villain
formulation of lattice Maxwell theory. The electric--magnetic duality was
discussed by relating the operators on the original lattice to those on the
dual lattice, which leads to the dual form of the Hamiltonian.  However, the
duality transformation itself was not formulated as a unitary operator acting
on the Hilbert space, and the theta term was not included.
To understand the full $\mathrm{SL}(2,\mathbb Z)$ structure of lattice Maxwell
theory in the Hamiltonian Villain formulation, several issues remain to be
clarified.  First, it is not yet clear how the $\mathcal S$ and $\mathcal T$ transformations should be realized as
operators on the Villain Hilbert space.  The $\mathcal S$ transformation is
expected to exchange electric and magnetic variables, while the
$\mathcal T$ transformation should implement the shift of the theta angle.
A natural candidate for constructing the $\mathcal{T}$ transformation is to use the Villain lattice
Chern--Simons action\footnote{See also \cite{Ikeda:2026lyl,Han:2021wsx,Han:2022cnr} for related works.} \cite{Jacobson:2023cmr,Jacobson:2024hov,Xu:2024hyo,DeMarco:2019pqv,Peng:2025nfa}, but this action has a staggered structure.
As a consequence, the magnetic contribution associated with the theta term is
naturally decomposed into staggered components rather than appearing as a
single Villain magnetic term.
This split structure makes the $\mathrm{SL}(2,\mathbb Z)$ analysis subtle.
In particular, it obscures how the Hamiltonian with the theta term should transform
under electric--magnetic duality, and it is not manifest whether the Villain
form of the Hamiltonian is preserved after the $\mathcal{S}$ transformation.  Thus, even if
one can identify lattice ingredients for the $\mathcal S$ and
$\mathcal T$ transformations, it remains nontrivial to show that they combine
into the expected modular structure at the Hamiltonian level.
Another subtlety arises from the fact that Chern--Simons theory at odd level is
a spin theory \cite{Dijkgraaf:1989pz,Belov:2005ze}. On the lattice, this is reflected  in the non-invariance of the
odd-level Chern--Simons action under the $\mathbb{Z}$ $1$-form gauge
transformation \cite{Jacobson:2023cmr}. Consequently, the $\mathcal{T}$ transformation obtained by
exponentiating the level-one Chern--Simons action does not preserve
the $\mathbb{Z}$ $1$-form gauge symmetry in general, whereas the $\mathcal{T}^{2}$
transformation constructed from the level-two Chern--Simons action does. To construct the $\mathcal{T}$ transformation, it would presumably be necessary
to incorporate fermionic degrees of freedom, as in \cite{Xu:2024hyo,Onoda:2025gqa}.

In this paper, we construct the $\mathcal S$ and $\mathcal T^{2}$
transformations in lattice Maxwell theory in the Hamiltonian Villain
formalism, and investigate the theta subgroup structure of $\mathrm{SL}(2,\mathbb{Z})$ generated by these
transformations.
We first review the Hilbert space, gauge structure, and constraints of the lattice Maxwell theory without the theta term discussed in \cite{Fazza:2022fss}. 
We then define an operator implementing the $\mathcal S$ transformation and show that, at $\theta=0$, the Hamiltonian is mapped to the same form with the coupling constant transformed as expected.
Next, we introduce the theta term and $\mathcal{T}^{2}$ transformation through the Euclidean topological charge \cite{Aoki:2026pvq,Jacobson:2023cmr,Anosova:2022cjm}. The theta term obtained in this way agrees with the construction of \cite{Lu:2026jnq}, and the $\mathcal{T}^{2}$ transformation shifts the theta angle by $4\pi$.
A characteristic feature of the lattice Hamiltonian is that the magnetic contribution to the theta term appears in a split form, reflecting the staggered structure of lattice Chern--Simons theory.
Although this prevents a direct $\mathcal S$ transformation from preserving the original Hamiltonian, the obstruction is associated with the non-zero momentum modes rather than with the zero momentum modes.
To make this structure explicit, 
we employ the Hodge decomposition to separate the zero momentum sector and pass to the momentum space.
After removing the non-zero momentum part by an appropriate unitary
transformation, the theta dependence is seen to reside entirely in the
zero momentum modes, while the non-zero momentum modes become independent of
theta. 
This allows us to define a modified $\mathcal S$ transformation operator that correctly reproduces the transformation \eqref{eq:coupling transformation} 
for the Hamiltonian with the theta term. 
In this sense, the lattice Hamiltonian retains the expected theta subgroup structure once the zero momentum sectors are isolated appropriately. 

We also extend the analysis to the case in which electric and magnetic charges are present.
In the Hamiltonian formulation, these charges are introduced as violations of
the Gauss law and the Bianchi identity, respectively. The Hilbert space
is decomposed into sectors labeled by fixed charge configurations
$\rho^{e}$ and $\rho^{m}$.  
After appropriately modifying the $\mathcal S$ and $\mathcal T^2$ transformations to incorporate the presence of charges and the framing effect, we find that the electric and magnetic charges transform as follows
\begin{align}
\label{eq:charge transformation by S flip}
\mathcal{S}:\quad (\rho^{e}_{x},\rho^{m}_{x,123}) &\;\longrightarrow\; (\rho^{m}_{x,123},-\rho^{e}_{x}),
\\
\mathcal{T}^{2}:\quad (\rho^{e}_{x},\rho^{m}_{x,123}) &\;\longrightarrow\; (\rho^{e}_{x}+2\rho^{m}_{x,123},\rho^{m}_{x,123}).
\end{align}
This establishes the expected theta subgroup structure also in
the presence of electric and magnetic charges. In particular, the transformation of the charges under $\mathcal T^2$ reproduces the Witten effect \cite{Witten:1979ey}.

We finally discuss a related non-invertible defect obtained by gauging the
$\mathbb{Z}_{N}$ subgroup of the global $\mathrm{U}(1)$ $1$-form symmetry.
Non-invertible symmetries have recently attracted much attention as
generalized symmetries described by topological defects whose fusion rules are
not group-like
\cite{Choi:2021kmx,Kaidi:2021xfk,Niro:2022ctq,Shao:2023gho}, and their lattice realizations have also been actively studied \cite{Aasen:2016dop,Choi:2021kmx,Koide:2021zxj}.
Maxwell theory is known to be self-dual under the gauging of a
$\mathbb{Z}_{N}$ global $1$-form symmetry, and the corresponding half-space
gauging gives rise to a non-invertible defect
\cite{Choi:2021kmx,Niro:2022ctq}.  In the Villain Hamiltonian formalism, we
realize this construction as an operator associated with gauging the
$\mathbb{Z}_{N}$ subgroup of the global $\mathrm{U}(1)$ $1$-form symmetry.  This
operator is closely related to the $\mathcal S$ transformation and can be
viewed as the Hamiltonian counterpart of the defect discussed in
\cite{Choi:2021kmx}. We then compute its fusion rule and show that the result reproduces the expected Tambara--Yamagami fusion rule \cite{Tambara:1998vmj}.

The organization of this paper is as follows. 
In Sec.~\ref{sec:Maxwell throty}, we review the Villain Hamiltonian formulation of lattice Maxwell theory, its Hilbert space, and gauge symmetries. 
In Sec.~\ref{sec:SL(2,Z)}, we construct the $\mathcal S$ and $\mathcal T^{2}$ transformations, study their action on the Hamiltonian with and without the theta term, and demonstrate the resulting theta subgroup structure of the lattice Maxwell theory. In Sec.~\ref{sec:with charge}, we modify the $\mathcal{S}$ and
$\mathcal{T}^{2}$ transformations so that the correct
theta subgroup structure is realized in the presence of electric
and magnetic charges.
In Sec.~\ref{sec:non-invertible}, we construct a non-invertible defect associated with gauging the
global $\mathbb{Z}_{N}$ $1$-form symmetry and study its fusion rule.
Finally, we conclude with a summary and discussion of future directions.
To make this paper self-contained, Appendices~\ref{app:Differential form} and~\ref{app:cup product} provide a brief review of
lattice differential forms and cup products. 
In addition, in Appendix~\ref{appendix:Unitarity of S}, we present a technical calculation showing the unitarity of the $\mathcal{S}$ transformation with the theta term and charges.

\section{Lattice Maxwell theory in Villain Hamiltonian formalism}
\label{sec:Maxwell throty}
In this section, we review the Villain Hamiltonian formulation of lattice Maxwell theory based on \cite{Fazza:2022fss}.
We consider a $3$-dimensional hypercubic lattice $\Lambda$, and denote lattice sites by $x = (x_{1},x_{2},x_{3})$ where $x_{i}\in \mathbb{Z}$.
We assume that the lattice forms a torus, with the identification $x_i \sim x_i + L$ for $i=1,2,3$.
An $r$-cell is defined as an oriented elementary hypercube of dimension $r$. In particular, $0$-cells correspond to lattice sites, $1$-cells to links, and $2$-cells to plaquettes. An $r$-cell is specified by a lattice site $x$ together with an ordered set of directions $\mu_{1}\mu_{2}\cdots\mu_{r}$ ($\mu_{1}<\mu_{2}<\cdots<\mu_{r}$), and is denoted by $(x,\mu_{1}\mu_{2}\cdots\mu_{r})$. 
An $r$-form is defined on the $r$-cell $(x,\mu_{1}\mu_{2}\cdots\mu_{r})$ of the lattice, whose value is assigned by $\alpha_{x,\mu_{1}\mu_{2}\cdots\mu_{r}}$. We assume that the $r$-form field is antisymmetric with respect to permutations of the direction indices.
The dual lattice $\tilde{\Lambda}$ is defined by shifting the original lattice $\Lambda$ by half a lattice spacing in each direction. More explicitly, the sites of $\tilde{\Lambda}$ are given by
$\tilde{x}=x+\frac{1}{2}\hat{s}$, where $\hat{s}=\hat{1}+\hat{2}+\hat{3}$,  $x\in \Lambda$, and $\hat{\mu}$ denotes the unit vector in the $\mu$-direction. We assume that the dual lattice $\tilde{\Lambda}$ also forms a $3$-dimensional torus with the same periodicity as the lattice $\Lambda$. For each oriented $r$-cell $(x,\mu_{1}\mu_{2}\cdots\mu_{r})$ on $\Lambda$, we define the dual cell as the oriented $(3-r)$-cell on the dual lattice $\tilde{\Lambda}$, denoted by $*(x,\mu_{1}\mu_{2}\cdots\mu_{r})$.
Let $\nu_{1}<\nu_{2}<\cdots<\nu_{3-r}$ be the complementary directions determined by
\begin{align}
\{\mu_{1},\mu_{2},\cdots,\mu_{r}\}\sqcup
\{\nu_{1},\nu_{2},\cdots,\nu_{3-r}\}
=
\{1,2,3\}.
\end{align}
Then the dual cell is defined by
\begin{align}
*(x,\mu_{1}\mu_{2}\cdots\mu_{r})
=
(\tilde{x}-\hat{\nu}_{1}-\hat{\nu}_{2}-\cdots-\hat{\nu}_{3-r},\nu_{1}\nu_{2}\cdots\nu_{3-r}).
\end{align}
The duality map $*$ gives a one-to-one correspondence between $r$-cells on $\Lambda$ and $(3-r)$-cells on $\tilde{\Lambda}$. Having defined the $r$-cells of the dual lattice, we can also introduce $r$-forms on the dual lattice.

Let us define the Hamiltonian for the lattice Maxwell theory in the
Villain formulation \cite{Villain:1974ir}. In this formulation, the electromagnetic field is represented by an $\mathbb{R}$-valued $1$-form operator $\hat{A}^e$ and $\mathbb{Z}$-valued $2$-form $\hat{n}$. The Villain Hamiltonian  of the lattice Maxwell theory without a theta term is given by \cite{Fazza:2022fss}

\begin{align}
\label{eq:Maxwell Hamiltonian without theta}
\hat{H}=\frac{1}{2\beta}
\sum_{(x,\mu)}
\left(\hat{\Pi}^{e}_{x,\mu}\right)^{2}+
\frac{\beta}{2}
\sum_{(x,\mu\nu)}
\left(
d\hat{A}^{e}+
2\pi \hat{n}\right)_{x,\mu\nu}^{2},
\end{align}
where $\beta$ is the inverse coupling constant, and $\hat{\Pi}^e_{x,\mu}$ is the canonical conjugate operator to $\hat{A}^e_{x,\mu}$. 
Here $d$ denotes the lattice exterior derivative. We refer the reader to Appendix~\ref{app:Differential form} for details on the lattice differential form notation.
These operators satisfy the commutation relation
\begin{align}
\label{commutation relation A}
\left[ \hat{A}^e_{x,\mu}, \hat{\Pi}^e_{x',\mu'} \right]
= i \, \delta_{x,x'} \, \delta_{\mu,\mu'}
= i \, \delta_{(x,\mu),(x',\mu')}.
\end{align}
We also define $\hat{A}^m_{*(x,\mu\nu)}$ as the canonical conjugate operator to $\hat{n}_{x,\mu\nu}$. These operators satisfy
\begin{align}
\left[ \hat{A}^m_{*(x,\mu\nu)}, \hat{n}_{x',\mu'\nu'} \right]
= i \, \delta_{(x,\mu\nu),(x',\mu'\nu')}.
\end{align}

We define the Hilbert space as the tensor product of local Hilbert spaces associated with links and plaquettes. The link Hilbert space is spanned by eigenstates of the operator $\hat{A}^e_{x,\mu}$, while the plaquette Hilbert space is spanned by eigenstates of the operator $\hat{n}_{x,\mu\nu}$. Thus, a basis of the total Hilbert space is given by the simultaneous eigenstates
\begin{align}
\ket{\{A^e\},\{n\}}
\equiv
\bigotimes_{(x,\mu)} \ket{A^e_{x,\mu}}
\otimes
\bigotimes_{(x,\mu\nu)} \ket{n_{x,\mu\nu}},
\end{align}
which satisfy
\begin{align}
\hat{A}^e_{x,\mu}\ket{\{A^e\},\{n\}}
&=
A^e_{x,\mu}\ket{\{A^e\},\{n\}},\\
\hat{n}_{x,\mu\nu}\ket{\{A^e\},\{n\}}
&=
n_{x,\mu\nu}\ket{\{A^e\},\{n\}}.
\end{align}
Here, $A^e_{x,\mu}\in\mathbb{R}$ and $n_{x,\mu\nu}\in\mathbb{Z}$ denote the eigenvalues of the corresponding operators.
The physical Hilbert space is restricted by the Gauss law constraint and no-monopole condition, which are given by
\begin{align}
(\partial \hat{\Pi}^{e})_{x}\ket{\mathrm{phys}}=0,\\
\label{eq:no-monopole}
(d\hat{n})_{x,\mu\nu\rho}\ket{\mathrm{phys}}=0.
\end{align}
For the definition of boundary operator $\partial$, see Appendix~\ref{app:Differential form}.

Next, we review the symmetries of the Hamiltonian \eqref{eq:Maxwell Hamiltonian without theta}.
The Hamiltonian enjoys several important symmetries. 
First, it is invariant under the ordinary $0$-form gauge symmetry,
\begin{align}
\label{eq:0 form gauge}
\hat{A}^{e}_{x,\mu}
\;\longrightarrow\;
\hat{A}^{e}_{x,\mu} + (d\lambda)_{x,\mu},
\end{align}
where $\lambda$ is an $\mathbb{R}$-valued $0$-form field. 
This is the standard gauge redundancy associated with the vector potential and is generated by the Gauss law operator $(\partial \hat{\Pi}^{e})_{x}$.
In addition, the Hamiltonian has a $\mathbb{Z}$ $1$-form gauge symmetry,
\begin{align}
\label{eq:Z 1-form gauge symmetry}
\hat{A}^{e}_{x,\mu}
&\;\longrightarrow\;
\hat{A}^{e}_{x,\mu}+2\pi k_{x,\mu},\\
\hat{n}_{x,\mu\nu}
&\;\longrightarrow\;
\hat{n}_{x,\mu\nu}-(dk)_{x,\mu\nu},
\end{align}
where $k_{x,\mu}$ is a $\mathbb{Z}$-valued $1$-form field. 
This symmetry reflects the compactness of the gauge field and ensures that the Villain variable $\hat{n}_{x,\mu\nu}$ compensates the discrete shift of $\hat{A}^{e}_{x,\mu}$ so that the gauge-invariant field strength $\hat{F}=d\hat{A}+2\pi \hat{n}$ remains unchanged.
The $\mathbb{Z}$ $1$-form gauge transformation operator is given by
\begin{align}
\label{eq:Z 1-form gauge transformation operator}
\hat{U}[k]=
\exp\left({
i\sum_{(x,\mu)} 2\pi k_{x,\mu}\hat{\Pi}^{e}_{x,\mu}
+i\sum_{(x,\mu\nu)} (dk)_{x,\mu\nu}\hat{A}^{m}_{*(x,\mu\nu)}}\right).
\end{align}
Since $k_{x,\mu}$ takes arbitrary integer values, gauge invariance under this transformation implies that the operator
\begin{align}
\label{eq:def m}
\hat{m}_{*(x,\mu\nu)}
=
\hat{\Pi}^{e}_{*(x,\mu\nu)}
+\frac{1}{2\pi}(d\hat{A}^{m})_{*(x,\mu\nu)}
\end{align}
must have integer eigenvalues. In other words, the physical states satisfy
\begin{align}
\label{eq:gauge invariance Z 1-form}
\hat{U}[k]\ket{\mathrm{phys}}=\ket{\mathrm{phys}} .
\end{align}
Furthermore, the Hamiltonian has a global $\mathbb{R}$ $1$-form symmetry,
\begin{align}
\label{eq:R 1-form symmetry}
\hat{A}^{e}_{x,\mu}
\;\longrightarrow\;
\hat{A}^{e}_{x,\mu}+\alpha_{x,\mu},
\end{align}
where $\alpha$ is an $\mathbb{R}$-valued $1$-form satisfying
$
d\alpha = 0.
$
Unlike the $\mathbb{Z}$ $1$-form gauge symmetry, this transformation is not a gauge redundancy but a genuine global symmetry. 
Physically, it corresponds to shifting the gauge field by a flat background $1$-form, and it acts nontrivially on line operators such as Wilson loops. Taking the quotient of the global $\mathbb{R}$ $1$-form symmetry by the global subgroup of the $\mathbb{Z}$ $1$-form gauge symmetry yields a global $\mathrm{U}(1)$ $1$-form symmetry, which is identified with the electric $\mathrm{U}(1)$ $1$-form symmetry of continuum Maxwell theory \cite{Gaiotto:2014kfa}.

\section{Theta subgroup structure of lattice Maxwell theory}
\label{sec:SL(2,Z)}
In this section, we study the theta subgroup structure of lattice Maxwell
theory in the Hamiltonian Villain formulation in the absence of electric and
magnetic charges.
We first construct the
$\mathcal S$ transformation at $\theta=0$ and show that it maps the
Hamiltonian to the same form with the expected transformation of the coupling.
We then define the theta term and $\mathcal T^{2}$ transformation by using the Euclidean topological charge \cite{Aoki:2026pvq,Jacobson:2023cmr,Anosova:2022cjm} and show that $\mathcal T^{2}$ transformation implements the $4\pi$
shift of the theta angle.
Although the direct $\mathcal S$ transformation does not preserve the original
form of the Hamiltonian in the presence of the theta term, we show that the
obstruction comes from the nonzero-momentum sector.  
To make this structure explicit, we perform the Hodge decomposition of the Villain field and
separate the harmonic sector. We then apply a unitary
transformation which removes the nonzero-momentum contribution of the theta
term. This allows us to define the appropriate $\mathcal S$ transformation in the presence
of the theta term and to show that, together with $\mathcal T^{2}$, it realizes
the theta subgroup of $\mathrm{SL}(2,\mathbb Z)$.

\subsection{$\mathcal{S}$ transformation}
It is known that the Maxwell theory is self-dual under the $\mathcal{S}$ transformation.
We first briefly recall the $\mathcal{S}$ transformation in the continuum Maxwell
theory.
The action of Maxwell theory with topological charge is given as
\begin{align}
S[A]=\frac{\beta}{2}\int F\wedge*F+\frac{i\theta}{8\pi^2}\int F\wedge F. \label{eq: Maxwell action}
\end{align}
Introducing the complex coupling constant
$
\tau
=
\frac{\theta}{2\pi}
+
2\pi i\beta ,
$
the $\mathcal{S}$ transformation acts as
\begin{align}
\label{eq:tau S transformation in section3}
\mathcal{S}:\tau \mapsto -\frac{1}{\tau}.
\end{align}
This transformation exchanges the electric and magnetic descriptions of the
theory and locally it can be derived by treating the field strength $F$ as an independent
$2$-form field, imposing the Bianchi identity by a Lagrange multiplier, and
integrating out $F$ \cite{Thompson:1995ks,Meynet:2025zem}.

The construction of \cite{Choi:2021kmx} gauges a global
$\mathbb{Z}_{N}$ $1$-form symmetry only on one side of an interface. Equivalently,
one introduces a $\mathbb{Z}_{N}$ gauge field and performs a Poisson resummation
with respect to this field only in half of spacetime. For the $\mathcal{S}$ transformation, the Villain variables already possess a
$\mathbb{Z}$ $1$-form gauge symmetry. Applying the same idea to this
gauge symmetry gives an interface operator, which is naturally identified with the
defect implementing the $\mathcal{S}$ transformation.

Motivated by this observation, we define the $\mathcal{S}$ transformation in the Hamiltonian formalism as follows
\begin{align}
\label{eq:S transformation}
\hat{\mathcal{S}}\ket{\{A^e\},\{n\}}
=\frac{1}{R}\int DA^{e\prime}\sum_{\{n'\}}\mathcal K[A^{e \prime},n';A^{e},n]\ket{\{A^{e\prime}\},\{n'\}},
\end{align}
where the kernel $\mathcal K[A^{e \prime },n';A^{e},n]$ is defined as
\begin{align}
\label{eq:kernel without theta}
\mathcal K[A^{e \prime },n';A^{e},n]=\exp\left[{\frac{i}{2\pi}\sum\left(A^{e\prime}\cup (dA^e+2\pi n)+2\pi n'\cup A^e \right)
}\right]
\end{align}
Here, $A^{e\prime}$ and $n'$ are, respectively, an $\mathbb{R}$-valued $1$-form and a $\mathbb{Z}$-valued $2$-form, in the same way as $A^e$ and $n$. $R$ is a normalization constant, which will be discussed later.  For the definition of the cup product, see Appendix \ref{app:cup product}.
The integration measure $\int \mathcal{D}A^{e\prime}$ and $\sum_{n'}$ are defined by
\begin{align}
\label{eq:measure}
\int \mathcal{D}A^{e\prime}
&=
\int_{-\pi}^{\pi}
\prod_{(x,\mu)} dA^{e\prime}_{x,\mu},\\
\sum_{\{n'\}}&=\prod_{(x,\mu\nu)}\sum_{n'_{x,\mu\nu}\in\mathbb{Z}}.
\end{align}
Under the $\mathbb{Z}$ $1$-form and $\mathbb{R}$ $0$-form gauge transformation for $A^{e}$ and $n$, \eqref{eq:kernel without theta} transforms as 
\begin{align}
&\mathcal K[A^{e \prime },n';A^{e}+2\pi k,n-dk]/\mathcal K[A^{e \prime },n';A^{e},n]=\exp\Bigr[2\pi i\sum n'\cup k \Bigl]=1,\\
&\mathcal K[A^{e \prime },n';A^{e}+d\lambda,n]/\mathcal K[A^{e \prime },n';A^{e},n]=\exp\Bigr[ i\sum n'\cup d\lambda \Bigl]=\exp\Bigr[ -i\sum dn'\cup \lambda \Bigl]=1.
\end{align}
In the last equality, we used the Leibniz rule for the cup product \eqref{eq:cup-leibniz} and the fact that $dn'=0$ holds in the physical Hilbert space.
Therefore, this kernel is gauge invariant.
Note that the following identity holds 
\begin{align}
A^{e\prime}\cup (dA^e+2\pi n)+2\pi n'\cup A^e =(dA^{e\prime}+2\pi n')\cup A^{e}+A^{e\prime}\cup2\pi n.
\end{align}
This form makes the gauge symmetry associated with the primed variables manifest.

Next, in order to see how the Hamiltonian transforms under the $\mathcal{S}$ transformation \eqref{eq:S transformation}, we first examine the transformation of the operator $\hat{\Pi}^{e}_{x,\mu}$. From the commutation relation, in the $A^{e}$ basis, the canonical momentum acts on the basis states as
\begin{align}
\hat{\Pi}^{e}_{x,\mu}\ket{\{A^e\}}
=
i\frac{\partial}{\partial A^{e}_{x,\mu}}\ket{\{A^e\}}.
\end{align}
Using this, we obtain
\begin{align}
\hat{\Pi}^{e\prime}_{x,\mu}\,&\hat{\mathcal S}\ket{\{A^e\},\{n\}}\nonumber\\
&=
\frac{1}{R}\int DA^{e\prime}\sum_{\{n'\}}
\mathcal K[A^{e \prime },n';A^{e},n]
\hat{\Pi}^{e\prime}_{x,\mu}\ket{\{A^{e\prime}\},\{n'\}}
\nonumber\\
&=
\frac{1}{R}\int DA^{e\prime}\sum_{\{n'\}}
\mathcal K[A^{e \prime },n';A^{e},n]
\left(
i\frac{\partial}{\partial A^{e\prime}_{x,\mu}}
\ket{\{A^{e\prime}\},\{n'\}}
\right)
\nonumber\\
&=
\frac{1}{R}\int DA^{e\prime}\sum_{\{n'\}}
\left(
-i\frac{\partial}{\partial A^{e\prime}_{x,\mu}}
\mathcal K[A^{e \prime },n';A^{e},n]
\right)
\ket{\{A^{e\prime}\},\{n'\}}
\nonumber\\
&=
\frac{1}{2\pi}(dA^e+2\pi n)_{x+\hat{\mu},\,\nu\rho}
\frac{1}{R}\int DA^{e\prime}\sum_{\{n'\}}
\mathcal K[A^{e \prime },n';A^{e},n]
\ket{\{A^{e\prime}\},\{n'\}}
\nonumber\\
&=
\hat{\mathcal S}\,
\frac{1}{2\pi}
\left(d\hat{A}^e+2\pi\hat{n}\right)_{x+\hat{\mu},\,\nu\rho}
\ket{\{A^e\},\{n\}}.
\end{align}
Similarly, one finds
\begin{align}
\hat{\mathcal S}&\hat{\Pi}^{e}_{x,\mu}\ket{\{A^e\},\{n\}}\nonumber\\
&=
\hat{\mathcal S}
\left(
i\frac{\partial}{\partial A^{e}_{x,\mu}}
\right)
\ket{\{A^e\},\{n\}}
\nonumber\\
&=
\frac{1}{R}\int DA^{e\prime}\sum_{\{n'\}}
\left(i\frac{\partial}{\partial A^{e}_{x,\mu}}\mathcal K[A^{e \prime },n';A^{e},n]
\right)
\ket{\{A^{e\prime}\},\{n'\}}
\nonumber\\
&=
\frac{1}{R}\int DA^{e\prime}\sum_{\{n'\}}
\left(-\frac{1}{2\pi}
\left(
dA^{e \prime}+2\pi n'
\right)_{x-\hat{\nu}-\hat{\rho},\,\nu\rho}\right)
\mathcal K[A^{e \prime },n';A^{e},n]
\ket{\{A^{e\prime}\},\{n'\}}
\nonumber\\
&=
-\frac{1}{2\pi}
\left(
d\hat{A}^{e \prime}+2\pi\hat{n}^{\prime}
\right)_{x-\hat{\nu}-\hat{\rho},\,\nu\rho}
\hat{\mathcal S}
\ket{\{A^e\},\{n\}}.
\end{align}
From the above calculation, we obtain
\begin{align}
\label{eq:S transformation for Pi}
\hat{\Pi}^{e\prime}_{x,\mu}\hat{\mathcal S}
&=
\hat{\mathcal S}\,
\frac{1}{2\pi}
\left(
d\hat{A}^e+2\pi\hat{n}
\right)_{x+\hat{\mu},\,\nu\rho},
\\
\hat{\mathcal S}\hat{\Pi}^{e}_{x,\mu}
&=
-\frac{1}{2\pi}
\left(
d\hat{A}^{e\prime}+2\pi\hat{n}^{\prime}
\right)_{x-\hat{\nu}-\hat{\rho},\,\nu\rho}
\hat{\mathcal S}.
\end{align}
Here, for each fixed $\mu$, the directions $\nu$ and $\rho$ are chosen such that
$\epsilon_{\mu\nu\rho}=1$, where $\epsilon_{\mu\nu\rho}$ denotes the three-dimensional Levi-Civita symbol. In what follows, whenever the indices $\mu$, $\nu$, and $\rho$ appear together, they are understood to be ordered according to the same convention.
From these relations, the Hamiltonian \eqref{eq:Maxwell Hamiltonian without theta} transforms as
\begin{align}
\hat{\mathcal S}\hat{H}=\hat{H}'\hat{\mathcal S},
\end{align}
where the transformed Hamiltonian $H'$ is given by
\begin{align}
\hat{H}'
&=
\frac{1}{2\beta(2\pi)^2}
\sum_{(x,\mu\nu)}
\left(d\hat{A}^{e\prime}+2\pi\hat{n}^{\prime}\right)_{x,\mu\nu}^{2}
+
\frac{\beta(2\pi)^2}{2}
\sum_{(x,\mu)}
\left(\hat{\Pi}^{e\prime}_{x,\mu}\right)^{2}
\\
&=
\frac{1}{2\beta'}
\sum_{(x,\mu)}
\left(\hat{\Pi}^{e\prime}_{x,\mu}\right)^{2}
+
\frac{\beta'}{2}
\sum_{(x,\mu\nu)}
\left(d\hat{A}^{e\prime}+2\pi\hat{n}'\right)_{x,\mu\nu}^{2}.
\end{align}
Here, $\beta'$ is defined as follows
\begin{align}
\beta'=\frac{1}{(2\pi)^2\beta}.
\end{align}
Thus, we find that the Hamiltonian is mapped to the same form under the $\mathcal S$ transformation, with the coupling constant transformed as $\beta\to\beta'$.
This transformation of the coupling constant is consistent with the modular transformation \eqref{eq:tau S transformation in section3}
of the complex coupling constant
$\tau$ at $\theta=0$.

Finally, we determine the normalization constant $R$ and demonstrate the unitarity of the $\mathcal{S}$ transformation.
Acting on the
basis state $\ket{\{A^{e}\},\{n\}}$, we obtain
\begin{align}
\hat{S}^{\dagger}\hat{S}\ket{\{A^{e}\},\{n\}}
&=
\frac{1}{R^{2}}
\int DA^{e\prime} DA^{e\prime\prime}
\sum_{\{n'\}}\sum_{\{n''\}}
\exp\left[
\frac{i}{2\pi}\sum
A^{e\prime}\cup
\left(
dA^{e}+2\pi n-dA^{e\prime\prime}-2\pi n''
\right)
\right]
\nonumber\\
&\hspace{3cm}\times
\exp\left[
i\, n'\cup (A^{e}-A^{e\prime\prime})
\right]
\ket{\{A^{e\prime\prime}\},\{n''\}},\nonumber\\
&=\frac{1}{R^{2}}
\int DA^{e\prime} DA^{e\prime\prime}
\sum_{\{n''\}}
\prod_{(x,\mu)}
\sum_{k_{x,\mu}\in\mathbb{Z}}
2\pi\,
\delta\!\left(A^{e}_{x,\mu}-A^{e\prime\prime}_{x,\mu}+2\pi k_{x,\mu}\right)
\nonumber\\
&\hspace{2cm}\times
\exp\left[
\frac{i}{2\pi}\sum
A^{e\prime}\cup
\left(
dA^{e}+2\pi n-dA^{e\prime\prime}-2\pi n''
\right)
\right]
\ket{\{A^{e\prime\prime}\},\{n''\}}.
\end{align}
Here we note that the integration over $A^{e\prime\prime}$ is restricted to the range
$[-\pi,\pi)$.  Performing the integral, the delta function selects a unique
integer $k$ satisfying 
\begin{align}
    A^{e\prime\prime} = A^{e} + 2\pi k .
\end{align}
Therefore, using the integer $k$ satisfying this relation, we can rewrite the
expression as follows
\begin{align}
\hat{S}^{\dagger}\hat{S}\ket{\{A^{e}\},\{n\}}
&=
\frac{(2\pi)^{3L^{3}}}{R^{2}}
\int DA^{e\prime}
\sum_{\{n''\}}
\exp\left[
\frac{i}{2\pi}\sum
A^{e\prime}\cup
\left(
2\pi n-2\pi n''-2\pi dk
\right)
\right]
\ket{\{A^{e}+2\pi k\},\{n''\}},\nonumber\\
&=\frac{(2\pi)^{3L^{3}}}{R^{2}}
\sum_{\{n''\}}2\pi\delta_{n-n''-dk,0}\ket{\{A^{e}+2\pi k\},\{n''\}},\nonumber\\
&=\frac{(2\pi)^{6L^{3}}}{R^{2}}\ket{\{A^{e}+2\pi k\},\{n-dk\}}.
\end{align}
Choosing the normalization
\begin{align}
R=(2\pi)^{3L^{3}},
\end{align}
we find
\begin{align}
\hat{S}^{\dagger}\hat{S}\ket{\{A^{e}\},\{n\}}
=
\ket{\{A^{e}+2\pi k\},\{n-dk\}}
=
\hat{U}[k]\ket{\{A^{e}\},\{n\}},
\end{align}
where $\hat{U}[k]$ is the operator implementing the $\mathbb{Z}$ 1-form gauge
transformation \eqref{eq:Z 1-form gauge transformation operator}.
Thus, on physical states, which are invariant under the $\mathbb{Z}$ 1-form
gauge transformation, we obtain
\begin{align}
\hat{S}^{\dagger}\hat{S}\ket{\mathrm{phys}}
=
\ket{\mathrm{phys}}.
\end{align}
Therefore, the $\mathcal{S}$ transformation is unitary on the physical Hilbert
space.

\subsection{$\mathcal{T}^{2}$ transformation}
In this section, we consider the $\mathcal{T}^{2}$ transformation.
We first recall the relation between the $4$-dimensional topological
charge and the $3$-dimensional lattice Chern--Simons action.
In the four-dimensional lattice Euclidean formulation, the topological charge $Q$  is given by \cite{Anosova:2022cjm,Aoki:2026pvq,Jacobson:2023cmr} 
\begin{align}
Q
&=
\sum \left[
\frac{1}{8\pi^{2}}
(dA^{e}+2\pi n)\cup(dA^{e}+2\pi n)
+
\frac{1}{4\pi}
(dA^{e}+2\pi n)\cup_{1}dn
\right] .
\label{eq:topological-charge}
\end{align}
Using the identities of the cup product and the higher cup product, this can
be decomposed as
\begin{align}
Q
&=
\sum
\frac{1}{8\pi^{2}}
d\Bigr[
A^{e}\cup dA^{e}
+2\pi\left(A^{e}\cup n+n\cup A^{e}\right)
+2\pi A^{e}\cup_{1}dn
\Bigl]
\nonumber\\
&\quad
+
\sum\left[
\frac{1}{2\pi}A^{e}\cup dn
+
\frac{1}{2}\left(n\cup n+n\cup_{1}dn\right)
\right] .
\label{eq:topological-charge-decomposition}
\end{align}
The first line is an exact term, whose boundary contribution gives the lattice Chern--Simons action.
 Therefore, when the $4$-dimensional
spacetime is cut along a $3$-dimensional hypersurface, this term induces
a $3$-dimensional Chern--Simons action on the hypersurface. The second
line is the bulk contribution. For the definition of $\cup_{1}$, see
Appendix~\ref{app:cup product}.
Following \cite{Jacobson:2023cmr}, we define the level $\kappa$ lattice
Chern--Simons action by\footnote{
In \cite{Jacobson:2023cmr}, the term $i\phi\cup dn$ is also included in the Chern--Simons action, where $\varphi$ is the $\mathbb{Z}$-valued $0$-form field acting as the Lagrange multiplier imposing $dn=0$. Here we omit this term, since the condition $dn=0$ is imposed directly as a
constraint on the physical Hilbert space.
}
\begin{align}
\label{eq:Chern-Simons action}
CS_{\kappa}
=
\sum
\Bigr[
\frac{\kappa}{4\pi}
\Bigl(
A^{e}\cup dA^{e} + 2\pi (A^{e}\cup n + n\cup A^{e})
\Bigr)
+ \frac{\kappa}{2}\, A^{e}\cup_{1} dn
\Bigr].
\end{align}
Let us now review the gauge structure of this Chern--Simons action.
Under the $\mathbb{Z}$ 1-form gauge transformation, the level one Chern--Simons action changes as
\begin{align}
\label{eq:Chern-Simons Z 1-form transformation}
&iCS_{1}({A}^{e}+2\pi k,{n}-dk)-iCS_{1}({A}^{e},{n})\nonumber\\
&=
\frac{i}{4\pi}
\sum
\Bigl[
2\pi k \cup (d{A}^{e}+2\pi {n})
-2\pi\, dk \cup {A}^{e}
-4\pi^{2}\, dk \cup k
+4\pi^{2}\, \hat{n}\cup k+4\pi^{2} k\cup_{1}d{n}
\Bigr]
\nonumber\\
&=
-\frac{i}{4\pi}
\sum d\!\left( 2\pi k \cup {A}^{e} \right)
+i\pi \sum \left( k\cup {n}+{n}\cup k-dk\cup k +k\cup_{1}d{n}\right).
\end{align}
The exact term vanishes after summing over the lattice.
The second term on the last line is an integer multiple of $i\pi$.
Therefore, after exponentiation, the transformation generally acquires an
overall sign. This is a manifestation of the fact that Chern--Simons theory at odd
level is a spin theory \cite{Dijkgraaf:1989pz,Belov:2005ze}.
On the other hand, for even level, the Chern--Simons action changes under a
$\mathbb Z$ $1$-form gauge transformation only by an integer multiple of
$2\pi i$.  Therefore, its exponentiated form is invariant under the
$\mathbb Z$ $1$-form gauge symmetry.
Under the $\mathbb{R}$ $0$-form gauge transformation, the level one Chern-Simons term changes as follows
\begin{align}
\label{eq:Chern-Simons R 0-form transformation}
&iCS_{1}({A}^{e}+d\lambda,{n})-iCS_{1}({A}^{e},{n})\nonumber\\
&=
\frac{i}{4\pi}\sum \Bigl[d\lambda \cup (d{A}^{e}+2\pi {n})+2\pi {n}\cup d\lambda +2\pi d\lambda \cup_{1}d{n}
\nonumber\Bigr]\\
&=
\frac{i}{4\pi}\sum \Bigl[d\!\left(\lambda \cup d{A}^{e}+\lambda \cup 2\pi {n}+2\pi {n}\cup \lambda\right)
-\lambda \cup 2\pi d{n}- 2\pi d{n}\cup \lambda +2\pi d\lambda \cup_{1}d{n}\Bigr]\nonumber \\
&=
\frac{i}{4\pi}\sum \Bigl[d\!\left(\lambda \cup d{A}^{e}+\lambda \cup 2\pi {n}+2\pi {n}\cup \lambda\right)
-2\lambda \cup 2\pi d{n}\Bigr] .
\end{align}
To obtain the last line, we used \eqref{eq:cup1-commutativity}.
The exact terms vanish after summing over the lattice.
The remaining terms contain $dn$, and in general they cannot be combined into
an exact term.  Thus, the Chern--Simons action is invariant under the
$\mathbb R$ $0$-form gauge symmetry only when the no-monopole condition
$dn=0$ is imposed.

In the Euclidean theory, shifting the theta angle changes the
action by the topological term. When spacetime is cut along a constant-time
slice, the exact part of this topological term gives a Chern--Simons action
on the spatial slice. Therefore, in the Hamiltonian formulation, the operator
implementing the $\mathcal{T}$ transformation is naturally obtained by exponentiating the lattice Chern--Simons action.
In the monopole-free sector of the Hamiltonian formulation, the constraint $d\hat{n}=0$ is imposed on physical states. Therefore, the terms \(A^{e}\cup_{1}dn\) are absent from the Chern-Simons action. We define the $\mathcal{T}$ transformation operator 
\begin{gather}
\hat{\mathcal{T}}
=
e^{
i{CS}_{1}(\hat{A}^{e},\hat{n})},
\\
{CS}_{1}(\hat{A}^{e},\hat{n})=\frac{1}{4\pi}\sum\Bigl[
\hat{A}^{e}\cup (d\hat{A}^{e}+2\pi \hat{n})+2\pi \hat{n}\cup \hat{A}^{e}\Bigr].
\end{gather}
The operator $\hat{\Pi}^{e}_{x,\mu}$ transforms under the $\mathcal{T}$ transformation as
\begin{align}
\hat{\mathcal{T}^{\dagger}}\,\hat{\Pi}^{e}_{x,\mu}\,\hat{\mathcal{T}}
=
\hat{\Pi}^{e}_{x,\mu}
+
\frac{1}{4\pi}
\left(d\hat{A}^{e}+2\pi \hat{n}\right)_{x-\hat{\nu}-\hat{\rho},\,\nu\rho}
+
\frac{1}{4\pi}
\left(d\hat{A}^{e}+2\pi \hat{n}\right)_{x+\hat{\mu},\,\nu\rho}.
\end{align}
From the fact that the $\mathcal {T}$ transformation shifts the theta angle by
$2\pi$, the Hamiltonian with the theta term in the monopole-free sector is
given as follows.\footnote{Unlike in the continuum Hamiltonian $
H=\int d^{3}x\,
\left[
\frac{1}{2\beta}
\left(
\mathbf{\Pi}
+
\frac{\theta}{4\pi^{2}}\mathbf{B}
\right)^{2}
+
\frac{\beta}{2}\mathbf{B}^{2}
\right]$, the theta term on lattice appears in a form in which the magnetic field is split into two parts. This reflects the fact that the lattice Chern-Simons action has a staggered symmetry \cite{Jacobson:2023cmr}.}
\begin{align}
\label{eq:Hamiltonian with theta}
\hat{H}(\beta,\theta)
&=
\frac{1}{2\beta}
\sum_{(x,\mu)}
\left(
\hat{\Pi}^{e}_{x,\mu}
+
\frac{\theta}{4\pi}
\left(
\frac{1}{2\pi}
(d\hat{A}^{\mathrm{e}}+2\pi \hat{n})_{x-\hat{\nu}-\hat{\rho},\,\nu\rho}
+
\frac{1}{2\pi}
(d\hat{A}^{\mathrm{e}}+2\pi \hat{n})_{x+\hat{\mu},\,\nu\rho}
\right)
\right)^{2}
\nonumber\\
&\qquad
+
\frac{\beta}{2}
\sum_{(x,\mu\nu)}
\left(
(d\hat{A}^{e}+2\pi \hat{n})_{x,\mu\nu}
\right)^{2}.
\end{align}
Note that although $\hat{\mathcal T}$ is not invariant under a $\mathbb Z$
$1$-form gauge transformation, the theta term itself is invariant under this
gauge transformation.
This way of introducing the theta term agrees with the construction given in \cite{Lu:2026jnq}.

Let us now examine more closely the relation between the $\mathcal T$
transformation and the generator of the $\mathbb Z$ $1$-form
gauge transformation $\hat{U}[k]$ defined in \eqref{eq:Z 1-form gauge transformation operator}.  The $\mathcal T$ transformation maps $\hat U[k]$ to a
twisted generator as
\begin{align}
\hat{\mathcal {T}^{\dagger}}\,\hat U[k]\hat{\mathcal{T}}
=
\hat V[k].
\end{align}
Using the transformation law of the level one lattice Chern--Simons action
under the $\mathbb Z$ $1$-form gauge transformation, we obtain
\begin{align}
\hat V[k]
&=
\exp\Bigr(
i\pi\sum
k\cup \hat n+\hat n\cup k-dk\cup k
\Bigl)
\hat U[k]\nonumber\\
&=
\exp\left({
i\sum_{(x,\mu)} 2\pi k_{x,\mu}\hat{\Pi}^{e}_{x,\mu}
+i\sum_{(x,\mu\nu)} (dk)_{x,\mu\nu}\hat{A}^{m}_{*(x,\mu\nu)}}+i\pi\sum (k\cup\hat{n}+\hat{n}\cup k)\right).
\end{align}
Thus, $\hat{\mathcal T}$ maps the ordinary
gauge generator $\hat U[k]$ to the twisted generator $\hat V[k]$. Accordingly, the condition imposed on the physical Hilbert space is changed
from invariance under $\hat U[k]$ to invariance under $\hat V[k]$. This modified gauge-invariance condition coincides with the modified Gauss-law
condition that appears in three-dimensional bosonization \cite{Chen:2018nog}. 
This indicates that, in order to construct a theory closed under the
$\mathcal{T}$ transformation, one has to introduce fermionic degrees of freedom,
as in \cite{Xu:2024hyo,Onoda:2025gqa}.

In the rest of this paper, we restrict ourselves to the bosonic theory and
consider the $\mathcal{T}^{2}$ transformation.
The operator $\hat{\mathcal{T}^{2}}$ commutes with $\hat{U}[k]$ and shifts the theta angle by $4\pi$,
\begin{align}
\hat{\mathcal{T}}^{2\dagger}\hat{H}(\beta,\theta)\hat{\mathcal{T}}^{2}=\hat{H}(\beta,\theta+4\pi).
\end{align}
In particular, the transformation of the complex coupling constant $\tau$ under the $\mathcal{T}^{2}$ transformation is $\tau\to\tau+2$,
and thus our result is consistent with this expectation.

\subsection{Theta subgroup structure}
\label{subsec:Theta sbgroup structure}
The $\mathcal{S}$ and $\mathcal{T}$ transformations are known to generate the
group structure called $\mathrm{SL}(2,\mathbb{Z})$. The subgroup of $\mathrm{SL}(2,\mathbb{Z})$ generated by
$\mathcal{S}$ and $\mathcal{T}^{2}$ is called the theta subgroup.
In this section, we consider the $\mathcal{S}$ transformation acting on the Hamiltonian with the theta term (\ref{eq:Hamiltonian with theta}), and verify that the $\mathcal{S}$ and $\mathcal{T}^{2}$ transformations form a theta subgroup of $\mathrm{SL}(2,\mathbb{Z})$. 
Let us first naively apply the $\mathcal S$ transformation defined in \eqref{eq:S transformation} to the Hamiltonian with a theta term. The Hamiltonian then transforms as follows,

\begin{align}
\hat{\mathcal{S}}\,\hat{H}(\beta,\theta)
=
\hat{H}'(\beta,\theta)\,\hat{\mathcal{S}},
\end{align}
where
\begin{align}
\hat{H}'(\beta,\theta)
&=
\frac{1}{2\beta}
\sum_{(x,\mu)}
\left(
-\frac{1}{2\pi}
\bigl(
d\hat{A}^{e'}+2\pi \hat{n}'
\bigr)_{x-\hat{\nu}-\hat{\rho},\,\nu\rho}
+
\frac{\theta}{4\pi}
\left(
\hat{\Pi}^{e'}_{x-\hat{s},\mu}
+
\hat{\Pi}^{e'}_{x,\mu}
\right)
\right)^{2}
\nonumber\\
&\qquad
+
\frac{\beta}{2}
\sum_{(x,\mu)}
\left(
2\pi \hat{\Pi}^{e'}_{x,\mu}
\right)^{2}.
\end{align}
Since $\hat{H}_{\theta}'$ contains products of $\hat{\Pi}^{e'}$ defined on different links, the Hamiltonian does not retain its original form after the direct $\mathcal{S}$ transformation. Therefore, the $\mathcal{S}$ transformation defined in \eqref{eq:S transformation} does not exhibit the correct transformation property in the presence of the theta term. 

Here, let us recall an important fact that the theta term contributes only to zero-momentum modes in the Hamiltonian \eqref{eq:Hamiltonian with theta} up to unitary transformation. 
Based on this fact, we modify the $S$ transformation defined in (\ref{eq:S transformation}) and show that the theta subgroup structure exists.
To show this fact explicitly, 
 we first perform the Hodge decomposition of $\hat{n}$ in the Hamiltonian with the theta term
\begin{align}
\hat{n} = d\hat{p} + \hat{h}.
\end{align}
Here, $\hat{h}$ satisfies $
d\hat{h}=0$ and $\partial \hat{h}=0.
$
Both $\hat{h}$ and $\hat{p}$ are $\mathbb{R}$-valued and $\hat{p}\in \rm{Im}(\partial^{(2)})$\footnote{
Since $n$ and $h$ are real, any complex solution $p=p_R+i p_I$ of
$n=dp+h$ satisfies $d p_I=0$. However, $p_I\in\rm{Im}(\partial^{(2)})$, and $\rm{Ker} (d^{(1)})\cap\operatorname{Im}(\partial^{(2)})={0}.$ Hence $p_I=0$, so $p$
is real.}. It is convenient to express the Hodge decomposition in terms of projection operators,
\begin{align}
P_{d}\hat{n}=d\hat{p},\qquad P_{0}\hat{n}=\hat{h}.
\end{align}
For details of the Hodge decomposition and the projection operators $P_{d}$ and $P_{0}$, see Appendix~\ref{app:Differential form}. 
Moreover, since $\hat{h}$ is a constant mode, it can be represented as
\begin{align}
\label{eq:h to n(0)}
\hat{h}_{\mu\nu}
=
\frac{1}{\sqrt{L^{3}}}
\hat{n}_{\mu\nu}(0),
\end{align}
where $\hat{n}_{\mu\nu}(0)$ is the zero momentum mode of $\hat{n}_{_{x,\mu\nu}}=\frac{1}{\sqrt{L^{3}}}\sum_{p}e^{ipx}\hat{n}_{\mu\nu}(p)$. 
Using this decomposition, the Hamiltonian can be rewritten as
\begin{align}
\hat{H}{(\beta,\theta)}
&=
\frac{1}{2\beta}
\sum_{(x,\mu)}
\left(
\hat{\Pi}^{e}_{x,\mu}
+
\frac{\theta}{4\pi}
\left[
\frac{1}{2\pi}
\bigl(d(\hat{A}^{e}+2\pi\hat{p})\bigr)_{x-\hat{\nu}-\hat{\rho},\,\nu\rho}
+
\frac{1}{2\pi}
\bigl(d(\hat{A}^{e}+2\pi\hat{p})\bigr)_{x+\hat{\mu},\,\nu\rho}
+
2\hat{h}_{\nu\rho}
\right]
\right)^{2}
\nonumber\\
&\qquad
+
\frac{\beta}{2}
\sum_{(x,\mu\nu)}
\left(
d\bigl(\hat{A}^{e}+2\pi\hat{p}\bigr)_{x,\mu\nu}
+
2\pi\hat{h}_{\mu\nu}
\right)^{2}.
\end{align} 
Next, let us consider the gauge-invariant unitary operator\footnote{
Note that under the $\mathbb{Z}$ $1$-form gauge transformation,
$\hat{A}^{e}\to \hat{A}^{e}+2\pi k$ and
$\hat{p}\to \hat{p}-P_{\partial}k$. Therefore,
$\hat{A}^{e}+2\pi \hat{p}$ transforms as
$\hat{A}^{e}+2\pi \hat{p}
\longrightarrow
\hat{A}^{e}+2\pi \hat{p}+2\pi(P_d+P_0)k.$
Since $(P_d+P_0)k$ is closed, the combination
$d(\hat{A}^{e}+2\pi \hat{p})$ is $\mathbb{Z}$ $1$-form gauge invariant.
Moreover, $\sum(\hat{A}^{e}+2\pi \hat{p})\cup d(\hat{A}^{e}+2\pi \hat{p})$ transforms as
\begin{align}
&\sum\bigl(\hat{A}^{e}+2\pi \hat{p}+2\pi(P_d+P_0)k\bigr)
\cup d(\hat{A}^{e}+2\pi \hat{p})
\nonumber\\
&=
\sum d\bigl(\hat{A}^{e}+2\pi \hat{p}+2\pi(P_d+P_0)k\bigr)
\cup (\hat{A}^{e}+2\pi \hat{p})
\nonumber\\
&=
\sum d(\hat{A}^{e}+2\pi \hat{p})\cup(\hat{A}^{e}+2\pi \hat{p})
\nonumber\\
&=
\sum(\hat{A}^{e}+2\pi \hat{p})\cup d(\hat{A}^{e}+2\pi \hat{p}).
\end{align}
Therefore, $\sum(\hat{A}^{e}+2\pi \hat{p})\cup d(\hat{A}^{e}+2\pi \hat{p})$ is also invariant under the
$\mathbb{Z}$ $1$-form gauge transformation.
}
\begin{align}
\label{eq:Unitary U}
\hat{U}
=
\exp\left(
-\frac{i\theta}{8\pi^{2}}
\sum
(\hat{A}^{e}+2\pi \hat{p})\cup d(\hat{A}^{e}+2\pi \hat{p})
\right).
\end{align}
By this unitary transformation, the $d(\hat{A}^{e}+2\pi\hat{p})$ dependent part in the
theta term can be removed. The transformed Hamiltonian is then given by
\begin{align}
{\hat{H}}^{\mathrm{red}}{(\beta,\theta)}
&=
\hat{U}^{\dagger}\hat{H}{(\beta,\theta)}\hat{U}
\nonumber\\
&=
\frac{1}{2\beta}
\sum_{(x,\mu)}
\left(
\hat{\Pi}^{e}_{x,\mu}
+
\frac{\theta}{2\pi}\hat{h}_{\nu\rho}
\right)^{2}
+
\frac{\beta}{2}
\sum_{(x,\mu\nu)}
\left(
d(\hat{A}^{e}+2\pi \hat{p})_{x,\mu\nu}
+
2\pi\hat{h}_{\mu\nu}
\right)^{2}.
\end{align}
The fact that the $d(\hat{A}^{e}+2\pi \hat{p})$ dependent part in the theta term can be
eliminated reflects that the theta term couples only to the zero momentum sector.

Next, we go to momentum space. We introduce the Fourier expansions
\begin{align}
\hat{A}^{e}_{x,\mu}
&=
\frac{1}{\sqrt{L^{3}}}
\sum_{p}
e^{ip\cdot x}\,
\hat{A}^{e}_{\mu}(p),
\\
\hat{\Pi}^{e}_{x,\mu}
&=
\frac{1}{\sqrt{L^{3}}}
\sum_{p}
e^{ip\cdot x}\,
\hat{\Pi}^{e}_{\mu}(p),
\\
\hat{p}_{x,\mu}
&=
\frac{1}{\sqrt{L^{3}}}
\sum_{p}
e^{ip\cdot x}\,
\hat{p}_{\mu}(p).
\end{align}
Here, the momentum $p=(p_{1},p_{2},p_{3})$ runs over the discrete Brillouin zone,
$
p_{\mu}=\frac{2\pi n_{\mu}}{L},
$
$
n_{\mu}=0,1,...,L-1.
$
The canonical commutation relations in momentum space are given by
\begin{align}
\bigl[\hat{A}^{e}_{\mu}(p),\hat{\Pi}^{e}_{\nu}(q)\bigr]
&=
i\,\delta_{\mu,\nu}\,\delta_{p+q,0}.
\end{align}
We also expand $\hat{n}_{x,\mu\nu}$ as
\begin{align}
\label{eq:Fourier for n}
\hat{n}_{x,\mu\nu}
=
\frac{1}{\sqrt{L^{3}}}
\sum_{p}
e^{ip\cdot x}\,
\hat{n}_{\mu\nu}(p).
\end{align}
As discussed above, the constant mode is proportional to the harmonic sector variable $\hat{h}_{\mu\nu}$ as in \eqref{eq:h to n(0)}.
Then, the reduced Hamiltonian takes the form
\begin{align}
\label{eq:reduced Hamiltonian}
\hat{H}^{\mathrm{red}}{(\beta,\theta)}
&=
\frac{1}{2\beta}
\sum_{p\neq 0}
\hat{\Pi}^{e}_{\mu}(p)\hat{\Pi}^{e}_{\mu}(-p)
\nonumber\\
&\qquad
+
\frac{\beta}{2}
\sum_{p\neq 0}
\Bigl(
f_{\mu}(p)(\hat{A}^{e}_{\nu}(p)+2\pi \hat{p}_{\nu}(p))
-
f_{\nu}(p)(\hat{A}^{e}_{\mu}(p)+2\pi\hat{p}_{\mu}(p))
\Bigr)\nonumber\\
&\qquad\qquad \times
\Bigl(
f_{\mu}(-p)(\hat{A}^{e}_{\nu}(-p)+2\pi\hat{p}_{\nu}(-p))
-
f_{\nu}(-p)(\hat{A}^{e}_{\mu}(-p)+2\pi\hat{p}_{\mu}(-p))
\Bigr)
\nonumber\\
&\qquad
+
\frac{1}{2\beta}
\left(
\hat{\Pi}^{e}_{\mu}(0)
+
\frac{\theta}{2\pi}\hat{n}_{\nu\rho}(0)
\right)^{2}
+
\frac{\beta}{2}
\bigl(
2\pi \hat{n}_{\mu\nu}(0)
\bigr)^{2},
\end{align}
where $f_{\mu}(p)=e^{ip\hat{\mu}}-1$.
This expression makes it manifest that the nonzero-momentum modes are independent of $\theta$, whereas the theta dependence appears only in the zero-mode.

We define the $\mathcal{S}$ transformation acting on the Hamiltonian with the theta term by
\begin{align}
\label{eq:def S theta in 3}
\hat{\mathcal S}_{\theta}\,
\bigl|\{A^{e}\}\{n\}\bigr\rangle
&=
\frac{1}{R_{\theta}}\int D A^{\prime e}
\sum_{\{n'\}}
\,\mathcal K_{\theta}[A^{e\prime},n';A^{e},n]\,
\bigl|\{A^{\prime e}\}\{n'\}\bigr\rangle ,
\end{align}
where the kernel \(\mathcal K_{\theta}[A',n';A,n]\) is given by
\begin{align}
\label{eq:kernel with theta}
\mathcal K_{\theta}[A^{e\prime},n';A^{e},n]
&=\exp\Biggl[i\sqrt{\beta\beta'}\sum \Bigr(A^{e\prime}\cup dA^{e}+A^{e\prime}\cup P_{d}2\pi n+P_{d}2\pi n'\cup A^{e}+P_{d}2\pi n'\cup2\pi p\Bigl)\Biggr]\nonumber\\
&\qquad\times\exp
\Bigl[\frac{i}{2\pi}\sum\Bigl(
A^{e\prime}\cup P_{0}2\pi n
+
P_{0}2\pi n'\cup A^{e}-P_{d}2\pi n'\cup2\pi p
\Bigl)
\Bigr]\nonumber\\
&=\exp\Biggl[
i\sqrt{\beta\beta'}
\sum_{p\neq 0}
A^{e\prime}_{\mu}(p)\,
e^{-ip_{\mu}}
\Bigl(
f_{\nu}(-p)A^{e}_{\rho}(-p)
-
f_{\rho}(-p)A^{e}_{\nu}(-p)
\Bigr)
\Biggr]
\nonumber\\
&\qquad\times\exp\Biggl[
i\sqrt{\beta\beta'}
\sum_{p\neq 0}
A^{e\prime}_{\mu}(p)\,
e^{-ip_{\mu}}
\Bigl(
f_{\nu}(-p)2\pi p_{\rho}(-p)
-
f_{\rho}(-p)2\pi p_{\nu}(-p)
\Bigr)
\Biggr]
\nonumber\\
&\qquad\times\exp\Biggl[
i\sqrt{\beta\beta'}
\sum_{p\neq 0}
2\pi p'_{\mu}(p)\,
e^{-ip_{\mu}}
\Bigl(
f_{\nu}(-p)A^{e}_{\rho}(-p)
-
f_{\rho}(-p)A^{e}_{\nu}(-p)
\Bigr)
\Biggr]
\nonumber\\
&\qquad\times\exp\Biggl[
i\sqrt{\beta\beta'}
\sum_{p\neq 0}
2\pi p'_{\mu}(p)\,
e^{-ip_{\mu}}
\Bigl(
f_{\nu}(-p)2\pi p_{\rho}(-p)
-
f_{\rho}(-p)2\pi p_{\nu}(-p)
\Bigr)
\Biggr]
\nonumber\\
&\qquad\times\exp\Biggl[
\frac{-i}{2\pi}
\sum_{p\neq 0}
2\pi p'_{\mu}(p)\,
e^{-ip_{\mu}}
\Bigl(
f_{\nu}(-p)2\pi p_{\rho}(-p)
-
f_{\rho}(-p)2\pi p_{\nu}(-p)
\Bigr)
\Biggr]
\nonumber\\
&\qquad\times
\exp
\Bigl[i\Bigl(
A^{e\prime}_{\mu}(0)\,n_{\nu\rho}(0)
+
n'_{\nu\rho}(0)\,A^{e}_{\mu}(0)
\Bigl)
\Bigr],
\end{align}
and $\beta'$ is defined as
\begin{align}
\label{eq:definition of beta'}
\beta'=\frac{\beta}{(2\pi\beta)^{2}+(\frac{\theta}{2\pi})^{2}}.
\end{align}
The overall factor $R_{\theta}$ acquires a dependence on $\theta$. We refer the reader to Appendix~\ref{appendix:Unitarity of S} for details.
Under the $\mathbb{Z}$ $1$-form gauge transformation acting on $A^{e}$ and $n$, the kernel transforms as follows
\begin{align}
\label{eq:kernel gauge transformation}
&\mathcal K_{\theta}[A^{e\prime},n';A^{e}+2\pi k,n-dk]/\mathcal K_{\theta}[A^{e\prime},n';A^{e},n]\nonumber\\
&=\exp\Biggl[i\sqrt{\beta\beta'}\sum \Bigr(P_{d}2\pi n'\cup 2\pi k-P_{d}2\pi n'\cup2\pi P_\partial k\Bigl)\Biggr]\nonumber\\
&\qquad\times\exp
\Bigl[\frac{i}{2\pi}\sum\Bigl(
P_{0}2\pi n'\cup 2\pi k+P_{d}2\pi n'\cup2\pi P_{\partial} k
\Bigl)
\Bigr]\nonumber\\
&=\exp\Biggl[i\sqrt{\beta\beta'}\sum \Bigr(2\pi p'\cup 2\pi dk-2\pi p'\cup2\pi dP_\partial k\Bigl)\Biggr]\nonumber\\
&\qquad\times\exp
\Bigl[\frac{i}{2\pi}\sum\Bigl(
P_{0}2\pi n'\cup 2\pi k+2\pi p'\cup2\pi dP_{\partial} k
\Bigl)
\Bigr]\nonumber\\
&=\exp\Biggl[i\sqrt{\beta\beta'}\sum \Bigr(2\pi p'\cup 2\pi dk-2\pi p'\cup2\pi dk\Bigl)\Biggr]\nonumber\\
&\qquad\times\exp
\Bigl[\frac{i}{2\pi}\sum\Bigl(
P_{0}2\pi n'\cup 2\pi k+2\pi p'\cup2\pi dk
\Bigl)
\Bigr]\nonumber\\
&=\exp
\Bigl[\frac{i}{2\pi}\sum\Bigl(
(P_{0}+P_{d})2\pi n'\cup 2\pi k
\Bigl)
\Bigr]=1.
\end{align}
In the sector without magnetic charges, the constraint $dn=dn'=0$ implies
that the coexact parts of $n$ and $n'$ vanish:
$
P_{\partial}n=P_{\partial}n'=0.
$ And we use the fact that $dP_{\partial}k=d(P_{d}+P_{\partial}+P_{0})k=dk$. The $\mathbb{Z}$ $1$-form gauge invariance with respect to $A^{e\prime}$ and $n'$ can be shown in the same way. Therefore, this kernel is gauge invariant.
Note that, at $\theta = 0$, $\sqrt{\beta\beta'}$ is equal to $\frac{1}{2\pi}$.
Therefore, at $\theta=0$,
\begin{align}
\mathcal K_{\theta=0}[A^{e\prime},n';A^{e},n]
&=\exp\Biggl[\frac{i}{2\pi}\sum (A^{e\prime}\cup dA^{e}+A^{e\prime}\cup (P_{d}+P_{0})2\pi n+(P_{d}+P_{0})2\pi n'\cup A^{e})\Biggr].
\end{align}
Thus, at $\theta=0$ \eqref{eq:kernel with theta} is equal to \eqref{eq:kernel without theta}.

The operator $\hat{\mathcal S}_{\theta}$ satisfies the following transformation properties
\begin{gather}
\hat{\mathcal S}_{\theta}\,\hat{\Pi}^{e}_{\mu}(p)
=
-
\sqrt{\beta\beta'}
\,e^{-ip_{\mu}}
\left(
f_{\nu}(p)(\hat{A}^{e\prime}_{\rho}(p)+2\pi\hat{p}_{\rho}(p))
-
f_{\rho}(p)(\hat{A}^{e\prime}_{\nu}(p)+2\pi\hat{p}_{\nu}(p))
\right)
\hat{\mathcal S}_{\theta},
\\
\hat{\mathcal S}_{\theta}\sqrt{\beta\beta'}\,
e^{ip_{\mu}}\left(
f_{\nu}(p)(\hat{A}^{e}_{\rho}(p)+2\pi\hat{p}_{\rho}(p))
-
f_{\rho}(p)(\hat{A}^{e}_{\nu}(p)+2\pi\hat{p}_{\nu}(p))
\right)
=
\hat{\Pi}^{e\prime}_{\mu}(p)\,
\hat{\mathcal S}_{\theta},
\\
\hat{\mathcal S}_{\theta}\,\hat{\Pi}^{e}_{\mu}(0)
=
-
\hat{n}'_{\nu\rho}(0)\,
\hat{\mathcal S}_{\theta},\\
\hat{\mathcal S}_{\theta}\,\hat{n}_{\nu\rho}(0)
=
\hat{\Pi}^{e\prime}_{\mu}(0)\,
\hat{\mathcal S}_{\theta}.
\end{gather}
These transformation properties imply that $\hat{\mathcal S}_{\theta}$ transforms the Hamiltonian as
\begin{align}
\hat{\mathcal S}_{\theta}\,\hat{H}^{\mathrm{red}}{(\beta,\theta)}
=
\hat{H}^{\prime\,\mathrm{red}}{(\beta',\theta')}\,
\hat{\mathcal S}_{\theta},
\end{align}
where
\begin{align}
\label{eq:transformed reduced Hamiltonian}
\hat{H}^{\mathrm{red}\prime}{(\beta',\theta')}
&=
\frac{1}{2\beta'}
\sum_{p\neq 0}
\hat{\Pi}^{e\prime}_{\mu}(p)\hat{\Pi}^{e\prime}_{\mu}(-p)
\nonumber\\
&\qquad
+
\frac{\beta'}{2}
\sum_{p\neq 0}
\left(
f_{\mu}(p)(\hat{A}^{e\prime}_{\nu}(p)+2\pi\hat{p}'_{\nu}(p))
-
f_{\nu}(p)(\hat{A}^{e\prime}_{\mu}(p)+2\pi\hat{p}'_{\mu}(p))
\right)\nonumber\\
&\qquad\qquad \times
\left(
f_{\mu}(-p)(\hat{A}^{e\prime}_{\nu}(-p)+2\pi\hat{p}'_{\nu}(-p))
-
f_{\nu}(-p)(\hat{A}^{e\prime}_{\mu}(-p)+2\pi\hat{p}'_{\mu}(-p))
\right)
\nonumber\\
&\qquad
+
\frac{1}{2\beta'}
\left(
\hat{\Pi}^{e\prime}_{\mu}(0)
+
\frac{\theta'}{2\pi}\hat{n}'_{\nu\rho}(0)
\right)^{2}
+
\frac{\beta'}{2}
\left(
2\pi \hat{n}'_{\mu\nu}(0)
\right)^{2}.
\end{align}
Here,
\begin{align}
\label{eq:beta theta transformed}
\beta'
=
\frac{\beta}{(2\pi\beta)^{2}+\left(\frac{\theta}{2\pi}\right)^{2}},
\qquad
\theta'
=
\frac{-\theta}{(2\pi\beta)^{2}+\left(\frac{\theta}{2\pi}\right)^{2}}.
\end{align}
Comparing \eqref{eq:reduced Hamiltonian} and \eqref{eq:transformed reduced Hamiltonian}, we find that
$\hat{\mathcal S}_{\theta}$ reproduces the transformation of $\tau$ under the
$\mathcal S$ transformation \eqref{eq:tau S transformation in section3}.
Therefore, $\hat{\mathcal S}_{\theta}$ is correctly defined as the operator
implementing the $\mathcal{S}$ transformation for the Hamiltonian with the theta term.

\section{Theta subgroup structure with electric and magnetic charges}
\label{sec:with charge}
In this section, we extend the analysis of the theta subgroup structure to sectors with
electric and magnetic charges.  These charges are introduced as violations of
the Gauss law and the Bianchi identity.  We show that the $\mathcal S$
transformation exchanges the two charge sectors, while the $\mathcal T^{2}$
transformation shifts the electric charge by twice the magnetic charge,
realizing the lattice Hamiltonian version of the Witten effect.  After
separating the dynamical and charge-sector variables by the Hodge
decomposition, we find that the dynamical Hamiltonian retains the same theta
subgroup structure as in the charge-free case.  We also discuss the lattice
shift appearing under $\mathcal S$ and interpret it as a framing flip.

\subsection{Transition of the electric and magnetic charges under $\mathcal{S}$ and $\mathcal{T}^{2}$ transformation in the absence of the theta term}
\label{subsec:charge without theta}
In this section, we discuss how the violations of the Gauss law and the Bianchi identity transform under the $\mathcal{S}$ and $\mathcal{T}^{2}$ transformations in the absence of the theta term.
In the Hamiltonian formulation, electric and magnetic charges are introduced as violations of the Gauss law and the Bianchi identity, respectively. 
The full Hilbert space is then decomposed as a direct sum over charge sectors,
\begin{align}
\mathcal{H}
=
\bigoplus_{\rho^{e},\rho^{m}}
\mathcal{H}_{\rho^{e},\rho^{m}}.
\end{align}
The Hilbert space $\mathcal{H}_{\rho^{e},\rho^{m}}$ is defined as the sector with fixed electric and magnetic charge configurations. Namely, physical states
$\ket{\rm{phys}}_{\rho^{e},\rho^{m}}\in \mathcal{H}_{\rho^{e},\rho^{m}}$
satisfy
\begin{align}
(\partial \hat{\Pi^{e}})_{x}\ket{\rm{phys}}_{\rho^{e},\rho^{m}}
&= \rho^{e}_{x}\ket{\rm{phys}}_{\rho^{e},\rho^{m}} \\
(d n)_{x,123}\ket{\rm{phys}}_{\rho^{e},\rho^{m}} &= \rho^{m}_{x,123}\ket{\rm{phys}}_{\rho^{e},\rho^{m}},
\end{align}
where $\rho^{\,e}_{x}$ is a $\mathbb{Z}$-valued $0$-form fixed electric charge density configuration and $\rho^{\,m}_{x,123}$ is a $\mathbb{Z}$-valued $3$-form fixed magnetic charge density configuration. 

\begin{figure}
    \centering
    \includegraphics[clip,width = 15.0cm]{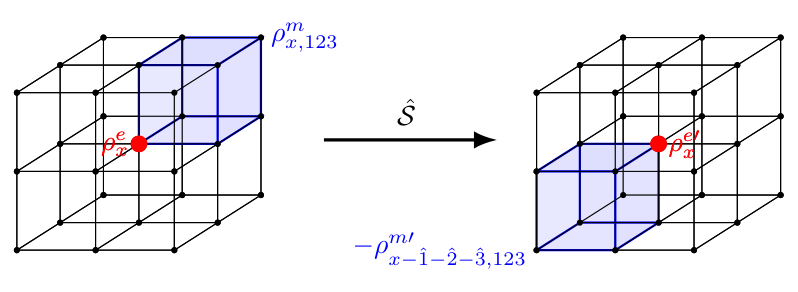}
    \caption{
    Illustration of the charge transformation.  The red point
    represents the electric charge $\rho^{e}_{x}$, while
    the blue cube represents the magnetic charge $\rho^{m}_{x,123}$.  Under the $\hat{\mathcal S}$ transformation, the
    charge characterized by $\rho^{e}_{x}$ and $\rho^{m}_{x,123}$ is mapped
    to
    $-\rho^{m\prime}_{x-\hat{1}-\hat{2}-\hat{3},123}$ and $\rho^{e\prime}_{x}$.
    }
    \label{fig:chare transformation}
\end{figure}

In the following, we examine how these charge densities transform under the $\mathcal{S}$ and $\mathcal{T}^{2}$ transformations.
Using the transformation law of operators under the $\mathcal{S}$ transformation \eqref{eq:S transformation for Pi}, one finds that $\partial \hat{\Pi}^{e}$ and $d\hat{n}$ are exchanged as
\begin{align}
\,(\partial \hat{\Pi}^{e\prime})_{x}\,\hat{\mathcal{S}}
&=
\,\hat{\mathcal{S}}(d\hat{n})_{x,\,123},
\\
-\,(d\hat{n}')_{x-\hat{s},\,123}\,\hat{\mathcal{S}}
&=
\hat{\mathcal{S}}(\partial \hat{\Pi}^{e})_{x},
\end{align}
where $\hat{s}=\hat{1}+\hat{2}+\hat{3}$.
Therefore, if we define the electric and magnetic charge density in the dual theory by
$
(\partial \hat{\Pi}^{e\prime})_{x}
=
{\rho}^{\,e\prime}_{x}$ and $
(d\hat{n}^{\prime})_{x,123}
=
{\rho}^{\,m\prime}_{x,123},
$
the above relations imply
\begin{align}
{\rho}^{e\prime}_{x}\,\hat{\mathcal{S}}
&=
\hat{\mathcal{S}}{\rho}^{\,m}_{x,123},
\\
-{\rho}^{m\prime}_{x-\hat{s},123}\,\hat{\mathcal{S}}
&=
\hat{\mathcal{S}}{\rho}^{\,e}_{x}.
\end{align}
Hence, under the $\mathcal{S}$ transformation, the violation of the Gauss law is mapped to the violation of the Bianchi identity, up to a minus sign and a lattice shift, while the violation of the Bianchi identity is mapped to the violation of the Gauss law. In other words, the $\mathcal{S}$ transformation exchanges electric and magnetic charges, up to the sign and the lattice shift appearing in the second relation. This relation is illustrated in Fig.~\ref{fig:chare transformation}. 

We next consider the $\mathcal{T}^{2}$ transformation. 
Since we no longer impose the no-monopole condition $d\hat{n}=0$, the Chern--Simons action entering the definition of the $\mathcal{T}^{2}$ transformation must include the $A\cup_{1}dn$ in \eqref{eq:Chern-Simons action} as well. 
Throughout this paper, $\hat{\mathcal T}^{2}$ denotes the $\mathcal {T}^{2}$ transformation operator defined in the no-monopole sector. 
When the no-monopole condition is relaxed, the corresponding $\mathcal{T}^{2}$ transformation operator is defined as 
\begin{gather}
\hat{\mathcal{T}}^{2}_{m}
=e^{iCS_{2,m}(\hat{A}^{e},\hat{n})}
\\
\label{eq:Chern-Simons term with cup1 in 4}
CS_{2,m}=
\sum
\left[
\frac{1}{2\pi}
\Bigl(
\hat{A}^{e}\cup d\hat{A}^{e}
+
2\pi \bigl(\hat{A}^{e}\cup \hat{n}+\hat{n}\cup \hat{A}^{e}\bigr)
\Bigr)
+
\,\hat{A}^{e}\cup_{1} d\hat{n}
\right].
\end{gather}
It follows from \eqref{eq:Chern-Simons Z 1-form transformation} that $\mathcal{T}^{2}_{m}$ is invariant under the
$\mathbb{Z}$ $1$-form gauge transformation.
On the other hand, $\mathcal{T}^{2}_{m}$ is not invariant under the
$\mathbb{R}$ $0$-form gauge transformation.  Instead, it transforms
$\partial\hat{\Pi}^{e}$ as follows
\begin{align}
\hat{\mathcal{T}}^{2\dagger}_{m}\,(\partial \hat{\Pi}^{e})_{x}\,\hat{\mathcal{T}}_{m}^{2}
=
(\partial \hat{\Pi}^{e})_{x}
+
2(d\hat{n})_{x,123}.
\end{align}
Therefore, the $\mathcal{T}^{2}$ transformation acts on the charge densities as
\begin{align}
\rho^{e}_{x}
&\;\longrightarrow\;
\rho^{e}_{x}+2\rho^{m}_{x,123},
\\
\rho^{m}_{x,123}
&\;\longrightarrow\;
\rho^{m}_{x,123}.
\end{align}
Thus, the $\mathcal{T}^{2}$ transformation shifts the electric charge by twice the magnetic charge, while leaving the magnetic charge unchanged.
This is the lattice Hamiltonian realization of the Witten effect \cite{Witten:1979ey}. 
As a result, a state carrying magnetic charge but no electric charge is transformed into a dyonic state under the $\mathcal{T}^{2}$ transformation. 
In terms of the charge vector, the two transformations act schematically as
\begin{align}
\label{eq:charge trasnformation by S in 4.1}
\mathcal{S}:\quad (\rho^{e}_{x},\rho^{m}_{x,123}) &\;\longrightarrow\; (\rho^{m}_{x,123},-\rho^{e}_{x+\hat{s}}),
\\
\label{eq:charge transformation by T in 4.1}
\mathcal{T}^{2}:\quad (\rho^{e}_{x},\rho^{m}_{x,123}) &\;\longrightarrow\; (\rho^{e}_{x}+2\rho^{m}_{x,123},\rho^{m}_{x,123}).
\end{align}
This agrees with the expected transformation of the charges under the theta
subgroup, except for the lattice shift appearing in the transformed $\rho^{m}$
under the $\mathcal{S}$ transformation.  We will
discuss this lattice shift in Section \ref{subsec:Flipping frames}.

\subsection{Theta subgroup structure of Hamiltonian with electric and magnetic charges}
In this section, we discuss the theta subgroup structure of the lattice Villain Hamiltonian with the theta term in the presence of electric and magnetic charges.
In the presence of electric and magnetic charges, the Hamiltonian with the theta term is written as follows
\begin{align}
\hat{H}{(\beta,\theta)}_{\rho_{e},\rho_{m}}
&=
\frac{1}{2\beta}
\sum_{(x,\mu)}
\Biggl(
\hat{\Pi}^{e}_{x,\mu}
+
\frac{\theta}{4\pi}
\Bigl[
\frac{1}{2\pi}
\left(
d\hat{A}^{e}
+
2\pi \hat{n}
\right)_{x-\hat{\nu}-\hat{\rho},\,\nu\rho}
\nonumber\\
&\hspace{3cm}
+
\frac{1}{2\pi}
\left(
d\hat{A}^{e}
+
2\pi \hat{n}
\right)_{x+\hat{\mu},\,\nu\rho}
+
(d\hat{n})_{x-g(\mu),\,123}
\Bigr]
\Biggr)^2
\nonumber\\
&\quad
+
\frac{\beta}{2}
\sum_{(x,\mu\nu)}
\left(
d\hat{A}^{e}
+
2\pi \hat{n}
\right)^2_{x,\mu\nu}.
\end{align}
Here, the shift function $g(\mu)$ is defined by
\begin{align}
g(1) = \hat{2} + \hat{3}, \quad
g(2) = \hat{3}, \quad
g(3) = 0.
\end{align}
The term $(d\hat{n})_{x-g(\mu),123}$ arises from the contribution of the $A \cup_1 d n$ term in the Chern--Simons action \eqref{eq:Chern-Simons term with cup1 in 4}. 
The action of $\hat{\mathcal T}^{2}_{m}$ shifts the theta parameter in this Hamiltonian as
$
\theta\rightarrow\theta + 4\pi
$, in other words $\tau\rightarrow\tau+2$.

To clarify the structure of the theta subgroup, we now decompose the Villain plaquette variable by the Hodge decomposition on the spatial lattice.  We write
\begin{align}
\hat n
=
d\hat p
+
\partial \hat q
+
\hat h,
\end{align}
where $d\hat p$ is the exact part, $\partial \hat q$ is the co-exact part, and
$\hat h$ is the harmonic part. In terms of projection operators they can be written as 
\begin{align}
P_{d}\hat{n}=d\hat{p},\qquad P_{\partial}\hat{n}=\partial\hat{p},\qquad P_{0}\hat{n}=\hat{h}.
\end{align}
We identify the magnetic charge part with the co-exact component,
\begin{align}
\label{eq:definition of n^{m}}
\hat{n}^{m} = P_{\partial}\hat{n}=\partial \hat q.
\end{align}
Indeed, using the nilpotency $d^{2}=0$ and the fact that $\hat{h}$ is harmonic, we find
$
d\hat{n}=d \hat{n}^{m}
=\rho^{m},
$
on the physical Hilbert space.
Note that under the $\mathbb{Z}$ 1-form gauge transformation, the operators transform as
\begin{align}
\hat A^{e} \;\longrightarrow\; \hat A^{e} + 2\pi k,
\qquad
\hat p \;\longrightarrow\; \hat p - P_{\partial}k ,
\end{align}
while $\hat n^{m}$ and $\hat{h}$ remain invariant.

With these decompositions, the Hamiltonian can be written as follows
\begin{align}
\hat{H}(\beta,\theta)_{\rho^{e},\rho^{m}}
&=
\frac{1}{2\beta}
\sum_{(x,\mu)}
\Biggl(
\hat{\Pi}^{e}_{x,\mu}
+
\frac{\theta}{4\pi}
\Bigl[
\frac{1}{2\pi}
\left(
d(\hat{A}^{e}+2\pi\hat{p})
+
2\pi \hat{n}^{m}+2\pi\hat{h}
\right)_{x-\hat{\nu}-\hat{\rho},\,\nu\rho}
\nonumber\\
&\hspace{3cm}
+
\frac{1}{2\pi}
\left(
d(\hat{A}^{e}+2\pi \hat{p})
+
2\pi \hat{n}^{m}+2\pi \hat{h}
\right)_{x+\hat{\mu},\,\nu\rho}
+
(d\hat{n}^{m})_{x-g(\mu),\,123}
\Bigr]
\Biggr)^2
\nonumber\\
&\quad
+
\frac{\beta}{2}
\sum_{(x,\mu\nu)}
\left(
d(\hat{A}^{e}+2\pi\hat{p})
+
2\pi \hat{n}^{m}+2\pi\hat{h}
\right)^2_{x,\mu\nu}.
\end{align}
We proceed in the same manner as section \ref{subsec:Theta sbgroup structure}.
We consider the unitary operator \eqref{eq:Unitary U}
which is gauge invariant.
Using this unitary operator, we remove the dependence on $(\hat{A}^{e}+2\pi\hat{p})$ appearing in the theta term.
Next, we focus on the term involving the magnetic component $\hat{n}^{m}$.
Using the identity of cup product \eqref{eq:cup1-commutativity}, one finds
\begin{align}
&(\hat{A}^{e}+2\pi\hat{p}) \cup \hat{n}^{m}
-
\hat{n}^{m} \cup (\hat{A}^{e}+2\pi\hat{p})\nonumber\\
&=
d\left((\hat{A}^{e}+2\pi\hat{p}) \cup_{1} \hat{n}^{m}\right)
-
d(\hat{A}^{e}+2\pi\hat{p}) \cup_{1} \hat{n}^{m}
+
(\hat{A}^{e}+2\pi\hat{p}) \cup_{1} d\hat{n}^{m},
\end{align}
we can rewrite the relevant terms.
Making use of this identity, we define another gauge invariant unitary operator\footnote{Under the $\mathbb{Z}$ $1$-form gauge transformation, neither $d(\hat{A}^{e}+2\pi\hat{p})$ nor
$\hat{n}^{m}$ transforms. Moreover, $d(\hat{A}^{e}+2\pi\hat{p})$ is invariant
under the $\mathbb{R}$ $0$-form gauge transformation.  Therefore, $\hat{V}$ is
gauge invariant.}
\begin{align}
\hat{V}
&=
\exp\left(
\frac{i\theta}{4\pi^{2}}
\sum
\left(
(\hat{A}^{e}+2\pi\hat{p}) \cup \hat{n}^{m}
-
\hat{n}^{m} \cup (\hat{A}^{e}+2\pi\hat{p})
-
(\hat{A}^{e}+2\pi\hat{p}) \cup_{1} d\hat{n}^{m}
\right)
\right),\nonumber\\
&=\exp\left(
-\frac{i\theta}{4\pi^{2}}
\sum
d(\hat{A}^{e}+2\pi\hat{p}) \cup_{1} \hat{n}^{m}
\right).
\end{align}
By performing the gauge-invariant unitary transformation,
the Hamiltonian is transformed as
\begin{align}
\label{eq:reduced Hamiltonian with charge}
\hat{H}^{\mathrm{red}}(\beta,\theta)_{\rho^{e},\rho^{m}}
&=
\hat{V}^{\dagger}\,\hat{U}^{\dagger}\,
\hat{H}(\beta,\theta)_{\rho^{e},\rho^{m}}
\hat{U}\hat{V},\nonumber\\
&=
\frac{1}{2\beta}
\sum_{(x,\mu)}
\Biggl(
\hat{\Pi}^{e}_{x,\mu}
+
\frac{\theta}{2\pi}
\hat{n}^{m}_{x+\hat{\mu},\,\nu\rho}
\nonumber
+
\frac{\theta}{2\pi}\hat{h}
_{\nu\rho}
\Biggr)^2
\nonumber\\
&\quad
+
\frac{\beta}{2}
\sum_{(x,\mu\nu)}
\left(
d(\hat{A}^{e}+2\pi \hat{p})
+
2\pi \hat{n}^{m}+2\pi\hat{h}
\right)^2_{x,\mu\nu}.
\end{align}

We now separate the contribution of the charges from the dynamical part of the
Hamiltonian.  We first decompose the electric momentum by the Hodge
decomposition on the spatial lattice.  For a lattice $1$-form $\hat\Pi^{e}$, we
write
\begin{align}
\hat\Pi^{e}
=
d\hat\phi
+
\partial\hat\chi
+
\hat\pi ,
\end{align}
where $d\hat\phi$ is the exact part, $\partial\hat\chi$ is the co-exact part,
and $\hat\pi$ is the harmonic part.  We identify the longitudinal and transverse part of $\Pi^{e}$ as 
\begin{align}
\hat\Pi^{e}_{\mathrm L}=P_{d}\hat{\Pi}^{e}
=
d\hat\phi,
\qquad \hat\Pi^{e}_{\mathrm T}
=(P_{\partial}+P_{0})\hat{\Pi}^{e}=
\partial\hat\chi+\hat\pi .
\end{align}
Then we have
\begin{align}
\partial\hat\Pi^{e}_{\mathrm L}=\rho^{e},
\qquad
\partial\hat\Pi^{e}_{\mathrm T}=0.
\end{align}
Thus $\hat\Pi^{e}_{\mathrm L}$ is fixed by the electric-charge density, while
$\hat\Pi^{e}_{\mathrm T}$ contains the transverse dynamical momentum.
We choose the representative of the magnetic-charge part as \eqref{eq:definition of n^{m}},
so that it satisfies
$
\partial\hat n^{m}=0.
$
Therefore it is orthogonal to the
exact part of $d(\hat A^{e}+2\pi\hat{p})$ in the magnetic term.  More explicitly, by using partial integration \eqref{eq:partial integration},
\begin{align}
\sum_{(x,\mu\nu)}
d(\hat A^{e}+2\pi\hat{p})_{x,\mu\nu}\,
\hat n^{m}_{x,\mu\nu}
=
\sum_{(x,\mu)}
(\hat{A}^{e}+2\pi\hat{p})_{x,\mu}\,
(\partial\hat n^{m})_{x,\mu}
=
0 .
\end{align}
Furthermore, the harmonic flux $\hat h$ is orthogonal to the co-exact part
$\hat n^{m}$
\begin{align}
\sum_{(x,\mu\nu)}
\hat h_{\mu\nu}\,\hat n^{m}_{x,\mu\nu}
=
0 .
\end{align}
Thus, the magnetic term in the Hamiltonian \eqref{eq:reduced Hamiltonian with charge} separates as
\begin{align}
&\frac{\beta}{2}
\sum_{(x,\mu\nu)}
\left(
d(\hat A^{e}+2\pi\hat{p})_{x,\mu\nu}
+
2\pi\hat n^{m}_{x,\mu\nu}
+
2\pi\hat h_{\mu\nu}
\right)^2
\nonumber\\
&=
\frac{\beta}{2}
\sum_{(x,\mu\nu)}
\left(
d(\hat A^{e}+2\pi\hat{p})_{x,\mu\nu}
+
2\pi\hat h_{\mu\nu}
\right)^2
+
\frac{\beta}{2}(2\pi)^2
\sum_{(x,\mu\nu)}
\left(
\hat n^{m}_{x,\mu\nu}
\right)^2 .
\end{align}
Next, we separate the momentum term.  Since
$\hat\Pi^{e}_{\mathrm T}$ is transverse, while
$\hat\Pi^{e}_{\mathrm L}$ is longitudinal, the two components are orthogonal
\begin{align}
\sum_{(x,\mu)}
\hat\Pi^{e}_{\mathrm T,x,\mu}
\hat\Pi^{e}_{\mathrm L,x,\mu}
=
0 .
\end{align}
Moreover, $\hat h$ belongs to the transverse sector, while
$\hat n^{m}$ pairs with the longitudinal sector.
Therefore the momentum term in \eqref{eq:reduced Hamiltonian with charge} separates as
\begin{align}
&\frac{1}{2\beta}
\sum_{(x,\mu)}
\Biggl(
\hat{\Pi}^{e}_{\mathrm T,x,\mu}
+
\hat{\Pi}^{e}_{\mathrm L,x,\mu}
+
\frac{\theta}{2\pi}
\hat{n}^{m}_{x+\hat{\mu},\,\nu\rho}
+
\frac{\theta}{2\pi}\hat{h}_{\nu\rho}
\Biggr)^2
\nonumber\\
&=
\frac{1}{2\beta}
\sum_{(x,\mu)}
\left(
\hat{\Pi}^{e}_{\mathrm T,x,\mu}
+
\frac{\theta}{2\pi}\hat{h}_{\nu\rho}
\right)^2
\nonumber\\
&\quad
+
\frac{1}{2\beta}
\sum_{(x,\mu)}
\left(
\hat{\Pi}^{e}_{\mathrm L,x,\mu}
+
\frac{\theta}{2\pi}
\hat n^{m}_{x+\hat\mu,\nu\rho}
\right)^2 .
\end{align}
Combining these results, the reduced Hamiltonian can be written as
\begin{align}
\hat H^{\mathrm{red}}{(\beta,\theta)_{\rho^e,\rho^m}}
=
\hat H^{\mathrm{dyn}}{(\beta,\theta)}
+
\hat H^{\mathrm{ch}}{(\beta,\theta)},
\end{align}
where
\begin{align}
\label{eq:dynical sector Hamiltonian}
\hat H^{\mathrm{dyn}}(\beta,\theta)
&=
\frac{1}{2\beta}
\sum_{(x,\mu)}
\left(
\hat{\Pi}^{e}_{\mathrm T,x,\mu}
+
\frac{\theta}{2\pi}\hat h_{\nu\rho}
\right)^2
+
\frac{\beta}{2}
\sum_{(x,\mu\nu)}
\left(
d(\hat A^{e}+2\pi\hat{p})_{x,\mu\nu}
+
2\pi\hat h_{\mu\nu}
\right)^2 ,
\\
\label{eq:charge sector Hamiltonian}
\hat H^{\mathrm{ch}}(\beta,\theta)
&=
\frac{1}{2\beta}
\sum_{(x,\mu)}
\left(
\hat{\Pi}^{e}_{\mathrm L,x,\mu}
+
\frac{\theta}{2\pi}
\hat n^{m}_{x+\hat\mu,\nu\rho}
\right)^2
+
\frac{\beta}{2}(2\pi)^2
\sum_{(x,\mu\nu)}
\left(
\hat n^{m}_{x,\mu\nu}
\right)^2 .
\end{align}

We now show that these Hamiltonians retain the theta subgroup structure. 
To make this statement precise, we go to the momentum space and recall the structure of the
transverse and longitudinal sector. 
As in position space, the structure of the transverse and the longitudinal sectors are conveniently written in terms of projection operators $P_{\partial}(p)$ and $P_{d}(p)$ for $p\neq0$ such that
\begin{align}
\hat\Pi^{e}_{\mathrm T,\mu}(p)
=
P_{\partial,\mu\nu}(p)\hat\Pi^{e}_{\nu}(p),\\
\hat\Pi^{e}_{\mathrm L,\mu}(p)
=
P_{d,\mu\nu}(p)\hat\Pi^{e}_{\nu}(p),
\end{align}
where the definitions of these projectors are given in Appendix~\ref{app:Differential form}, and we write them explicitly here as follows
\begin{align}
P_{\partial,\mu\nu}(p)
&=
\delta_{\mu\nu}
-
\frac{f_{\mu}(p)\,f_{\nu}(-p)}{|f(p)|^{2}},\\
P_{d,\mu\nu}(p)
&=
\frac{f_{\mu}(p)\,f_{\nu}(-p)}{|f(p)|^{2}}.
\end{align}
Using this projector, we define the transverse gauge field by
\begin{align}
\hat A^{e}_{\mathrm T,\mu}(p)
&=
\begin{cases}
P_{\partial,\mu\nu}(p)\hat A^{e}_{\nu}(p) & (p\neq0),\\
\hat A^{e}_{\mu}(0) & (p=0),
\end{cases}\\
\hat A^{e}_{\mathrm L,\mu}(p)
&=
\begin{cases}
P_{d,\mu\nu}(p)\hat A^{e}_{\nu}(p) & (p\neq0),\\
0 & (p=0).
\end{cases}
\end{align}
The canonical commutation relation for $p\neq0$ takes the form
\begin{align}
\left[
\hat A^{e}_{\mathrm T,\mu}(p),
\hat\Pi^{e}_{\mathrm T,\nu}(p')
\right]
&=
iP_{\partial,\mu\nu}(p)\delta_{p+p',0},\\
\left[
\hat A^{e}_{\mathrm L,\mu}(p),
\hat\Pi^{e}_{\mathrm L,\nu}(p')
\right]
&=
iP_{d,\mu\nu}(p)\delta_{p+p',0}.
\end{align}
Although the projector $P_{\partial}$ appears in the definition of the
transverse variables, it acts trivially on all physical combinations entering
the Hamiltonian and the duality kernel.  

We now define the $\mathcal S$ transformation as
\begin{align}
\label{eq:S with charge}
\hat{\mathcal S}^{\mathrm {T+L}}_{\theta}
\bigl|\{A^{e}\},\{n\}\bigr\rangle
&=
\frac{1}{R_{\theta}}\int DA^{e\prime}
\sum_{\{n'\}}
\mathcal K^{\mathrm T}_{\theta}
[A^{e\prime},n';A^{e},n]\mathcal K^{\rm{L}}
[A^{e\prime},n';A,n]\,
\bigl|\{A^{\prime e}\},\{n'\}\bigr\rangle .
\end{align}
The kernel for the transverse sector $\mathcal K^{\mathrm T}_{\theta}
[A^{e\prime},n';A^{e},n]$ is equal to the kernel defined in \eqref{eq:kernel with theta}
\begin{align}
\label{eq:kernel with theta transverse}
\mathcal K^{\mathrm T}_{\theta}
[A^{e\prime},n';A^{e},n]=\mathcal K_{\theta}
[A^{e\prime},n';A^{e},n].
\end{align} Note that, using the projection property \eqref{eq:cup and projection}, one can show that the kernel defined in \eqref{eq:kernel with theta} depends only on the transverse sector of $A^{e}$.
On the other hand, the kernel for the longitudinal sector is given as follows
\begin{align}
\label{eq:kernel longitudinal}
\mathcal K^{\rm{L}}
[A^{e\prime},n';A^{e},n]
&=
\exp\left[
\frac{i}{2\pi}\sum
\left(
A^{e\prime}\cup P_{\partial}2\pi n
+
P_{\partial}2\pi n'\cup A^{e}
\right)
\right],\nonumber\\
&=\exp\left[
\frac{i}{2\pi}\sum
\left(
A^{e\prime}_{\rm{L}}\cup P_{\partial}2\pi n
+
P_{\partial}2\pi n'\cup A^{e}_{\rm{L}}
\right)
\right],\nonumber\\
&=\exp\left[
\frac{i}{2\pi}\sum
\left(
A^{e\prime}_{\rm{L}}\cup 2\pi n^{m}
+
2\pi n^{m\prime}\cup A^{e}_{\rm{L}}
\right)
\right].
\end{align}
In the second line, we used the projection property \eqref{eq:cup and projection}.
We first examine the structure of gauge symmetry for these kernels. As we see in \eqref{eq:kernel gauge transformation}, the kernel for the transverse sector transforms under the $\mathbb{Z}$ $1$-form gauge transformation of $A^{e}$ and $n$ as
\begin{align}
\label{eq:kernel^T Z 1-form}
\mathcal K^{\rm{T}}_{\theta}[A^{e\prime},n';A^{e}+2\pi k,n-dk]/\mathcal K^{\rm{T}}_{\theta}[A^{e\prime},n';A^{e},n]=\exp
\Bigl[\frac{i}{2\pi}\sum\Bigl(
(P_{0}+P_{d})2\pi n'\cup 2\pi k
\Bigl)
\Bigr].
\end{align}
Since charges are now present, $P_{\partial}n'$ is generally nonzero, and hence this factor is not equal to unity in general. On the other hand, the $\mathbb{Z}$ $1$-form gauge transformation of the kernel for the longitudinal sector is given by
\begin{align}
\label{eq:kernel^L Z 1-form}
\mathcal K^{\rm{L}}[A^{e\prime},n';A^{e}+2\pi k,n-dk]/\mathcal K^{\rm{L}}[A^{e\prime},n';A^{e},n]=\exp
\Bigl[\frac{i}{2\pi}\sum\Bigl(
P_{\partial}2\pi n'\cup 2\pi k
\Bigl)
\Bigr].
\end{align}
Combining \eqref{eq:kernel^T Z 1-form} and \eqref{eq:kernel^L Z 1-form}, we find that the full kernel, obtained by combining the transverse and longitudinal sectors, is gauge invariant. This $\mathcal S$ transformation is unitary on the physical Hilbert space. We refer the reader to Appendix~\ref{appendix:Unitarity of S} for details.

The transformation laws for the transverse sector following from this $\mathcal{S}$ transformation are
\begin{gather}
\hat{\mathcal S}^{\rm{T}+\rm{L}}_{\theta}\,\hat{\Pi}^{e}_{\rm{T},\mu}(p)
=
-
\sqrt{\beta\beta'}
\,e^{-ip_{\mu}}
\left(
f_{\nu}(p)(\hat{A}^{e\prime}_{\rm{T},\rho}(p)+2\pi\hat{p}_{\rho}(p))
-
f_{\rho}(p)(\hat{A}^{e\prime}_{\rm{T},\nu}(p)+2\pi\hat{p}_{\nu}(p))
\right)
\hat{\mathcal S}^{\rm{T}+\rm{L}}_{\theta},
\\
\hat{\mathcal S}^{\rm{T}+\rm{L}}_{\theta}\sqrt{\beta\beta'}\,
e^{ip_{\mu}}\left(
f_{\nu}(p)(\hat{A}^{e}_{\rm{T},\rho}(p)+2\pi\hat{p}_{\rho}(p))
-
f_{\rho}(p)(\hat{A}^{e}_{\rm{T},\nu}(p)+2\pi\hat{p}_{\nu}(p))
\right)
=
\hat{\Pi}^{e\prime}_{\rm{T},\mu}(p)\,
\hat{\mathcal S}^{\rm{T}+\rm{L}}_{\theta},
\\
\hat{\mathcal S}^{\rm{T}+\rm{L}}_{\theta}\,\hat{\Pi}^{e}_{\rm{T},\mu}(0)
=
-
\hat{n}'_{\nu\rho}(0)\,
\hat{\mathcal S}^{\rm{T}+\rm{L}}_{\theta},\\
\hat{\mathcal S}^{\rm{T}+\rm{L}}_{\theta}\,\hat{n}_{\nu\rho}(0)
=
\hat{\Pi}^{e\prime}_{\rm{T},\mu}(0)\,
\hat{\mathcal S}^{\rm{T}+\rm{L}}_{\theta}.
\end{gather}
Alternatively, using the projection operators in position space, the relations can be written as follows
\begin{gather}
\hat{\mathcal{S}}^{\rm{T}+\rm{L}}_{\theta}P_{\partial}\hat\Pi^{e}_{x,\mu}=-\sqrt{\beta\beta'}(d\hat{A}^{e\prime}+P_{d}2\pi \hat{n}')_{x-\hat{\mu},\nu\rho}\hat{\mathcal{S}}^{\rm{T}+\rm{L}}_{\theta},\\
P_{\partial}\hat{\Pi}^{e\prime}_{x,\mu}\hat{\mathcal{S}}^{\rm{T}+\rm{L}}_{\theta}=\hat{\mathcal{S}}^{\rm{T}+\rm{L}}_{\theta}\sqrt{\beta\beta'}(d\hat{A}^{e}+P_{d}2\pi \hat{n})_{x+\hat{\mu},\nu\rho},
\\
\hat{\mathcal{S}}^{\rm{T}+\rm{L}}_{\theta}P_{0}\hat{\Pi}^{e}_{x,\mu}=-\frac{1}{2\pi}(P_{0}2\pi \hat{n}')_{x-\hat{\mu},\nu\rho}\hat{\mathcal{S}}^{\rm{T}+\rm{L}}_{\theta},\\
P_{0}\hat{\Pi}^{e\prime}_{x,\mu}\hat{\mathcal{S}}^{\rm{T}+\rm{L}}_{\theta}=\hat{\mathcal{S}}^{\rm{T}+\rm{L}}_{\theta}\frac{1}{2\pi}(P_{0}2\pi \hat{n})_{x+\hat{\mu},\nu\rho}.
\end{gather}
These relations imply that the dynamical sector of the Hamiltonian transforms as
\begin{align}
\hat{\mathcal S}^{\mathrm{T}+\rm{L}}_{\theta}\,
\hat H^{\mathrm{dyn}}(\beta,\theta)\,
=
\hat H^{\mathrm{dyn}\prime}(\beta',\theta')\,
\hat{\mathcal S}^{\mathrm{T}+\rm{L}}_{\theta},
\end{align}
where
\begin{align}
\hat H^{\mathrm{dyn}\prime}(\beta',\theta')
&=
\frac{1}{2\beta'}
\sum
\left(
\hat\Pi_{\mathrm T,x,\mu}^{\prime e}
+
\frac{\theta'}{2\pi}\hat h'_{\nu\rho}
\right)^2
+
\frac{\beta'}{2}
\sum
\left(
d\hat (A_{\mathrm T}^{e\prime}+2\pi\hat{p})_{x,\mu\nu}
+
2\pi\hat h'_{\mu\nu}
\right)^2 .
\end{align}
The transformed parameters are given in \eqref{eq:beta theta transformed}.
Equivalently, in terms of the complex coupling constant $\tau$, this is precisely
\begin{align}
\mathcal{S}:\tau\longrightarrow -\frac{1}{\tau}.
\end{align}
Thus, after separating the charge-sector variables, the transverse dynamical
part of the Hamiltonian has exactly the same theta subgroup
structure as in the charge-free theory. 
The transformation laws for the longitudinal sector following from the $\mathcal{S}$ transformation \eqref{eq:S with charge} are given as
\begin{align}
\hat{\Pi}_{\mathrm L,x,\mu}^{\prime e}\,
\hat{\mathcal S}^{\rm{T}+\rm{L}}_{\theta}
=
\hat{\mathcal S}^{\rm{T}+\rm{L}}_{\theta}\,
\hat n^{m}_{x+\hat\mu,\nu\rho},
\end{align}
\begin{align}
-\hat n^{\prime m}_{x-\hat\nu-\hat\rho,\nu\rho}\,
\hat{\mathcal S}^{\rm{T}+\rm{L}}_{\theta}
=
\hat{\mathcal S}^{\rm{T}+\rm{L}}_{\theta}\,
\hat{\Pi}^{e}_{\mathrm L,x,\mu}.
\end{align}
Taking the divergence of these relations, and using
$
\partial \hat{\Pi}^{e}_{\mathrm L}=\rho^{e},
$ and $
d\hat n^{m}=\rho^{m}
$ on the physical Hilbert space,
we obtain
\begin{align}
\rho^{\,e\prime}_{x}\,
\hat{\mathcal S}^{\rm{T}+\rm{L}}_{\theta}
&=
\hat{\mathcal S}^{\rm{T}+\rm{L}}_{\theta}\,
\rho^{\,m}_{x,123},
\\
-\rho^{\,m\prime}_{x-\hat s,123}\,
\hat{\mathcal S}^{\rm{T}+\rm{L}}_{\theta}
&=
\hat{\mathcal S}^{\rm{T}+\rm{L}}_{\theta}\,
\rho^{\,e}_{x}.
\end{align}
Thus, 
$\hat{\mathcal S}^{\rm{T}+\rm{L}}_{\theta}$ implements the electric-magnetic
exchange in the charge sector in the same way as in \eqref{eq:charge trasnformation by S in 4.1}.
Furthermore, under $\hat{\mathcal S}^{\rm{T}+\rm{L}}_{\theta}$, the charge-sector Hamiltonian transforms as
\begin{align}
\hat H^{\rm{ch\prime}}\,
\hat{\mathcal S}^{\rm{T}+\rm{L}}_{\theta}
=
\hat{\mathcal S}^{\rm{T}+\rm{L}}_{\theta}\,
\hat H^{\mathrm{ch}},
\end{align}
where the transformed Hamiltonian is given as
\begin{align}
\label{eq:H^{ch}'}
\hat H^{\mathrm{ch\prime}}
&=
\frac{1}{2\beta}
\sum_{(x,\mu)}
\left(
-\hat n^{m\prime}_{x-\hat\nu-\hat\rho,\nu\rho}
+
\frac{\theta}{2\pi}
\hat{\Pi}^{e\prime}_{\mathrm L,x,\mu}
\right)^2
+
\frac{\beta}{2}(2\pi)^2
\sum_{(x,\mu\nu)}
\left(
\hat{\Pi}^{e\prime}_{\mathrm L,x,\mu}
\right)^2
\nonumber\\
&=
\frac{1}{2\beta'}
\sum_{(x,\mu)}
\left(
\hat{\Pi}^{e\prime}_{\mathrm L,x,\mu}
+
\frac{\theta'}{2\pi}
\hat n^{m\prime}_{x-\hat\nu-\hat\rho,\nu\rho}
\right)^2
+
\frac{\beta'}{2}(2\pi)^2
\sum_{(x,\mu\nu)}
\left(
\hat n^{m\prime}_{x,\mu\nu}
\right)^2.
\end{align}
The transformed parameters $\beta'$ and $\theta'$ are the same as in \eqref{eq:beta theta transformed}.
The Hamiltonian retains the same quadratic structure as in \eqref{eq:charge sector Hamiltonian}, although the
theta-dependent coupling involving $\hat n^{m}$ is shifted by
$-\hat{s}=-\hat{1}-\hat{2}-\hat{3}$. 
This shift of $\hat{n}^{m}$ originates
from the point-split structure of the cup product, and is closely analogous to
the framing flip appearing in the analysis of Wilson loops in \cite{Aoki:2026pvq}. We discuss the flipping of the framing in the next section.

Let us consider the $\mathcal{S}$ transformation at $\theta=0$, combining the transverse and longitudinal parts. The combined $\mathcal{S}$ transformation becomes
\begin{align}
S^{\rm{T}+\rm{L}}_{\theta=0}\ket{\{A^{e}\},\{n\}}
&=
\frac{1}{R}\int DA^{e\prime}
\sum_{\{n'\}}\,
\mathcal K^{\rm{T}}_{\theta=0}
[A^{e\prime},n';A,n]\mathcal K^{\rm{L}}
[A^{e\prime},n';A,n]\,
\ket{\{A^{e\prime}\},\{n'\}},
\end{align}
where
\begin{align}
&\mathcal K^{\rm{T}}_{\theta=0}
[A^{e\prime},n';A,n]\mathcal K^{\rm{L}}
[A^{e\prime},n';A,n]\nonumber\\
&=\exp\Biggl[\frac{i}{2\pi}\sum (A^{e\prime}\cup dA^{e}+A^{e\prime}\cup P_{d}2\pi n+P_{d}2\pi n'\cup A^{e}+P_{d}2\pi n'\cup2\pi p)\Biggr]\nonumber\\
&\qquad\times\exp
\Bigl[\frac{i}{2\pi}\sum\Bigl(
A^{e\prime}\cup P_{0}2\pi n
+
P_{0}2\pi n'\cup A^{e}-P_{d}2\pi n'\cup2\pi p
\Bigl)
\Bigr]\nonumber\\
&\qquad\times
\exp\left[
\frac{i}{2\pi}\sum
\left(
A^{e\prime}\cup P_{\partial}2\pi n
+
P_{\partial}2\pi n'\cup A^{e}
\right)
\right]\nonumber\\
&=\exp[\frac{i}{2\pi}\sum (A^{e\prime}\cup dA^{e}+A^{e\prime}\cup (P_{d}+P_{\partial}+P_{0})2\pi n+(P_{d}+P_{\partial}+P_{0})2\pi n'\cup A^{e})]\nonumber\\
&=\exp[\frac{i}{2\pi}\sum (A^{e\prime}\cup dA^{e}+A^{e\prime}\cup 2\pi n+2\pi n'\cup A^{e})].
\end{align}
This kernel coincides with the one defined in \eqref{eq:kernel without theta}.

The $\mathcal {T}^{2}$ transformation can be treated in a completely analogous
manner to the discussion in Section \ref{subsec:charge without theta}. As a
result, one finds that $\hat{\mathcal{T}}_{m}^{2}$ shifts the electric charge by twice the magnetic charge \eqref{eq:charge transformation by T in 4.1},
while leaving the magnetic charge invariant and shifting theta by $4\pi$, in other words $\tau\rightarrow \tau+2$.

In summary, both the $\mathcal S$ and $\mathcal{T}^{2}$ transformations act on the dynamical part of the 
Hamiltonian and the charge sectors in a manner consistent with the expected
theta subgroup structure except for the framing flip.  The $\mathcal S$ transformation exchanges
electric and magnetic charges up to a lattice shift originating from the
point-split structure of the cup product, while the $\mathcal{T}^{2}$ transformation
shifts the electric charge by twice the magnetic charge and simultaneously shifts
$\theta$ by $4\pi$.

\subsection{Flipping the framing}
\label{subsec:Flipping frames}

We now discuss the flipping of the framing.  In the Euclidean modified Villain
formulation, it was pointed out in \cite{Aoki:2026pvq} that the naive
operation corresponding to the $\mathcal S$ transformation, namely the Poisson
summation, exchanges electric and magnetic variables but at the same time
flips the framing of dyonic line operators.  Equivalently, the Poisson
summation implements the electromagnetic exchange together with a shift of
the lattice position between the electric and magnetic parts.  In that
formulation, the framing flip can be characterized by 
$(\mathcal{PT})^{3}$, where $\mathcal{P}$ is the Poisson summation and $\mathcal{T}$ is the $\mathcal{T}$ transformation. 
A closely related effect appears in the Hamiltonian formulation.  As
shown above, the charge sector $\mathcal {S}$ transformation exchanges
the electric and magnetic charges, but the transformed magnetic charge contains
a lattice shift \eqref{eq:charge trasnformation by S in 4.1}.  We interpret this shift as the Hamiltonian
counterpart of the framing flip. 
To remove this shift, we modify the kernel for the longitudinal sector as follows\footnote{In the Hamiltonian formulation, we do not directly compute the operator
$(\mathcal{S}\mathcal{T})^{3}$ on the Hilbert space, and hence we do not
separate out the framing-flip effect in this way.  Instead, we construct an
$\mathcal{S}$ transformation which already includes the effect of the framing
flip by modifying the kernel obtained from the Poisson summation.}
\begin{align}
\mathcal{K}^{\rm{L}}[A^{e\prime},n';A^{e},n]\longrightarrow\mathcal K^{\rm{L}}_{\mathrm{fl}}
[A^{e\prime},n';A^{e},n]
\end{align}
where the modified kernel is given by
\begin{align}
\label{eq:kernel flipped}
\mathcal K^{\rm{L}}_{\mathrm{fl}}
[A^{e\prime},n';A^{e},n]
&=
\exp\left[
\frac{i}{2\pi}\sum
\left(
A^{e\prime}\cup P_{\partial}2\pi n
+
P_{\partial}2\pi n'\cup A^{e} +P_{d}A^{e}\cup_{1}dP_{\partial}2\pi n'-2\pi p\cup_{1}dP_{\partial}2\pi n'\right)
\right],\nonumber\\
&=\exp\left[
\frac{i}{2\pi} \sum
\left(
A^{e\prime}\cup P_{\partial}2\pi n
+
A^{e}\cup P_{\partial}2\pi n'-2\pi p\cup_{1}dP_{\partial}2\pi n'\right)
\right],\nonumber\\
&=\exp\left[
\frac{i}{2\pi} \sum
\left(
A^{e\prime}_{\rm{L}}\cup 2\pi n^{m}
+
A^{e}_{\rm{L}}\cup 2\pi n^{m\prime}-2\pi p\cup_{1}dn^{m\prime}\right)
\right].
\end{align}
We use the identity of the cup product given in \eqref{eq:cup1-commutativity}, explicitly, it is written as
\begin{align}
P_{d}A^{e}\cup P_{\partial}2\pi n'
-
P_{\partial}2\pi n'\cup P_{d}A^{e}
=
d(P_{d}A^{e}\cup_{1}P_{\partial}2\pi n')+
P_{d}A^{e}\cup_{1} dP_{\partial}2\pi n'
-
dP_{d}A^{e}\cup_{1} P_{\partial}2\pi n',
\end{align}
to obtain the second line.
Let us note that this modification does not spoil the gauge-symmetry structure of the $\mathcal S$ transformation. 
Indeed, under the $\mathbb{Z}$ 1-form transformation for the $A^{e}$ and $n$, the added term transforms as 
\begin{align}
&P_{d}A^{e}\cup_{1}dP_{\partial}2\pi n'-2\pi p\cup_{1}dP_{\partial}n'\nonumber\\
&=A^{e}\cup_{1}dP_{\partial}2\pi n'-P_{\partial}(A^{e}+2\pi p)\cup_{1}dP_{\partial}2\pi n'\nonumber\\
&\longrightarrow A^{e}\cup_{1}dP_{\partial}2\pi n'+2\pi k\cup_{1} dP_{\partial}2\pi n'-P_{\partial}(A^{e}+2\pi p+(P_{d}+P_{0})2\pi k)\cup_{1}dP_{\partial}2\pi n'\nonumber\\
&=A^{e}\cup_{1}dP_{\partial}2\pi n'+2\pi k\cup_{1} d2\pi n'-P_{\partial}(A^{e}+2\pi p)\cup_{1}dP_{\partial}2\pi n'
\end{align}
Since the variation in the exponent of \eqref{eq:kernel flipped} is an integer multiple of $2\pi i$, the added term does not spoil the structure of the $\mathbb{Z}$ $1$-form gauge symmetry. Moreover, under the $\mathbb{Z}$ 1-form transformation for the $A^{e\prime}$ and $n'$, the added term is invariant because $P_{\partial}2\pi n'$ does not transform. 

Using the $\mathcal{S}$ transformation operator $\hat{\mathcal{S}}^{\rm{T}+\rm{L}}_{\rm{fl}}$ obtained by using the modified longitudinal part of the kernel \eqref{eq:kernel flipped} and the transverse part of the kernel \eqref{eq:kernel with theta transverse}, we find
\begin{align}
\hat{\Pi}_{\mathrm L,x,\mu}^{e\prime}\,
\hat{\mathcal S}^{\rm{T}+\rm{L}}_{\mathrm{fl}}
&=
\hat{\mathcal S}^{\rm{T}+\rm{L}}_{\mathrm{fl}}\,
\hat n^{m}_{x,\nu\rho},
\\
-\hat n^{m\prime}_{x,\nu\rho}\,
\hat{\mathcal S}^{\rm{T}+\rm{T}}_{\mathrm{fl}}
&=
\hat{\mathcal S}^{\rm{T}+\rm{L}}_{\mathrm{fl}}\,
\hat{\Pi}_{\mathrm L,x,\mu}^{e}.
\end{align}
Furthermore, at the level of charge densities, we obtain
\begin{align}
\rho^{e\prime}_{x}\,
\hat{\mathcal S}^{\rm{T}+\rm{L}}_{\mathrm{fl}}
&=
\hat{\mathcal S}^{\rm{T}+\rm{L}}_{\mathrm{fl}}\,
\rho^{m}_{x,123},
\\
-\rho^{m\prime}_{x,123}\,
\hat{\mathcal S}^{\rm{T}+\rm{L}}_{\mathrm{fl}}
&=
\hat{\mathcal S}^{\rm{T}+\rm{L}}_{\mathrm{fl}}\,
\rho^{e}_{x},
\end{align}
on the physical Hilbert space.
Thus, the modified $\mathcal{S}$ transformation exchanges the electric and magnetic charges without any lattice shift.
\begin{align}
\mathcal{S}_{\mathrm{fl}}:\quad (\rho^{e}_{x},\rho^{m}_{x,123}) &\;\longrightarrow\; (\rho^{m}_{x,123},-\rho^{e}_{x}),
\end{align}
We next consider the transformation of the Hamiltonian in the charge sector.  
We find that
\begin{align}
\hat H^{\mathrm{ch}\prime}(\beta',\theta')\,
\hat{\mathcal S}^{\rm{T}+\rm{L}}_{\mathrm{fl}}
=
\hat{\mathcal S}^{\rm{T}+\rm{L}}_{\mathrm{fl}}\,
\hat H^{\mathrm{ch}}(\beta,\theta),
\end{align}
where
\begin{align}
\hat H^{\mathrm{ch}\prime}(\beta,\theta)
&=
\frac{1}{2\beta'}
\sum_{(x,\mu)}
\left(
\hat{\Pi}_{\mathrm L,x,\mu}^{e\prime}
+
\frac{\theta'}{2\pi}
\hat n^{m\prime}_{x+\hat{\mu},\nu\rho}
\right)^{2}
+
\frac{\beta'}{2}(2\pi)^{2}
\sum_{(x,\mu\nu)}
\left(
\hat n^{m\prime}_{x,\mu\nu}
\right)^{2}.
\end{align}
Therefore, the one-site shift of $\hat n^{m}$ that appeared in the Hamiltonian \eqref{eq:H^{ch}'} transformed by the original $\mathcal{S}$ transformation is removed by the framing-flip modification. Note that the transformation properties of the dynamical sector are unaffected by the modification associated with the framing flip.

\section{Gauging of $\mathbb{Z}_{N}$ 1-form symmetry and fusion rule}
\label{sec:non-invertible}
In this section, we study the non-invertible defect obtained by gauging the
$\mathbb{Z}_{N}$ subgroup of the global $\mathrm{U}(1)$ $1$-form symmetry in lattice Maxwell
theory. We first discuss the theory without the theta term and charges, and
show that the resulting defect obeys a Tambara--Yamagami fusion rule. We then extend the
construction to the theory with the theta term and electric and magnetic
charges. We show that the fusion rule is unchanged and that the non-invertible defect transforms the complex coupling constant as
$
\tau \longrightarrow -\frac{N^{2}}{\tau},
$
while preserving the form of the Hamiltonian.

\subsection{Non-invertible defect without the theta term and charges}
In this section, we discuss the defect arising from gauging the $\mathbb{Z}_{N}$ global $1$-form symmetry in the absence of the theta term and charges.
The lattice Maxwell theory has the global $\mathbb{R}$ $1$-form symmetry given in \eqref{eq:R 1-form symmetry}. Taking the quotient of this symmetry by the global subgroup of the $\mathbb{Z}$ $1$-form gauge symmetry yields a global $\mathrm{U}(1)$ $1$-form symmetry.
By gauging the $\mathbb{Z}_{N}$ subgroup of this global $1$-form symmetry, one obtains a defect for which Maxwell theory is known to be self-dual. 

In the lattice Maxwell theory, the defect associated with gauging the $\mathbb{Z}_{N}$ global $1$-form symmetry is defined as
\begin{align}
\label{eq:non-invertible defect}
\hat{\mathcal N}\ket{\{A^{e}\},\{n\}}
=
\frac{1}{R_{N}}
\int DA^{e\prime}\sum_{\{n'\}}
\exp\left[
\frac{iN}{2\pi}
\sum
\left(
A^{e\prime}\cup (dA^{e}+2\pi n)
+
2\pi n'\cup A^{e}
\right)
\right]
\ket{\{A^{e\prime}\},\{n'\}}.
\end{align}
The term in the exponent is precisely the boundary Chern--Simons coupling that appears in the Euclidean formulation of the lattice Villain Maxwell theory discussed in \cite{Choi:2021kmx}, when the spacetime manifold is cut into two regions and the $\mathbb{Z}_{N}$ global $1$-form symmetry is gauged on one side.
Note that the ${\mathcal S}$ transformation defined in \eqref{eq:S transformation} is the $N=1$ case of this defect.
The transformation of operators under this defect is given by
\begin{align}
\label{eq:N transformation for Pi}
\hat{\Pi}^{e\prime}_{x,\mu}\hat{\mathcal N}
&=
\hat{\mathcal N}\,
\frac{N}{2\pi}
\left(
d\hat{A}^e+2\pi\hat{n}
\right)_{x+\hat{\mu},\,\nu\rho},
\\
\label{eq:N transformation for Pi2}
\hat{\mathcal N}\hat{\Pi}^{e}_{x,\mu}
&=
-\frac{N}{2\pi}
\left(
d\hat{A}^{e\prime}+2\pi\hat{n}^{\prime}
\right)_{x-\hat{\nu}-\hat{\rho},\,\nu\rho}
\hat{\mathcal N}.
\end{align}
Next, we consider the fusion rule of the defect $\hat{\mathcal{N}}$. It is known that the following relations hold between the $\hat{\mathcal{N}}$ and the operator $\hat{\eta}$ generating the global $\mathbb{Z}_{N}$ $1$-form symmetry \cite{Choi:2021kmx,Kaidi:2021xfk}
\begin{align}
\label{eq:fusion rule eta N}
\hat{\mathcal N}\times \hat{\eta} = \hat{\mathcal N},\qquad
\hat{\eta}^{\prime}\times \hat{\mathcal N} = \hat{\mathcal N}.
\end{align}
Here, the generator of the global $\mathbb{Z}_{N}$ 1-form symmetry $\hat{\eta}$ is defined as
\begin{align}
\hat{\eta}[k]=\mathrm{exp}\left[{
\frac{2\pi i}{N}
\sum_{(x,\mu)}
k_{x,\mu}
\left(
\hat{\Pi}^{e}_{x,\mu}
+
\frac{1}{2\pi}
(d\hat{A}^{m})_{*(x,\mu)}
\right)}\right],
\end{align}
where $k_{x,\mu}$ is a $\mathbb{Z}$-valued $1$-form parameter satisfying $dk=0$. Note that, since $\hat{m}$ defined in \eqref{eq:def m} has integer eigenvalues,
one finds $\hat{\eta}^{N}=1$. By lattice integration by parts, we have
\begin{align}
\sum_{(x,\mu)}
k_{x,\mu}(d\hat{A}^{m})_{*(x,\mu)}
=
\sum_{(x,\mu\nu)}
(dk)_{x,\mu\nu}\hat{A}^{m}_{*(x,\mu\nu)}.
\end{align}
Since $dk=0$, $\hat{\eta}$ can be rewritten as
\begin{align}
\hat{\eta}[k]
=
\mathrm{exp}\left[{\frac{2\pi i}{N}
\sum_{(x,\mu)}
k_{x,\mu}\hat{\Pi}^{e}_{x,\mu}}\right].
\end{align}
From the transformation \eqref{eq:N transformation for Pi}, it follows that
\begin{align}
\hat{\mathcal N}\times \hat{\eta}
&=
\hat{\mathcal N}\mathrm{exp}\left[{\frac{2\pi i}{N}
\sum_{(x,\mu)}
k_{x,\mu}\hat{\Pi}^{e}_{x,\mu}}\right]\nonumber\\
&=\mathrm{exp}\left[{-i
\sum_{(x,\mu)}
k_{x,\mu}}
\left(
d\hat{A}^{e\prime}+2\pi\hat{n}^{\prime}
\right)_{x-\hat{\nu}-\hat{\rho},\,\nu\rho}\right]\hat{\mathcal{N}}\nonumber\\
&=\mathrm{exp}\left[-i
\sum
(d\hat{A}^{e\prime}+2\pi \hat{n}')\cup k \right]\hat{\mathcal{N}}.
\end{align}
Since $\sum dA^{e\prime}\cup k=\sum A^{e\prime}\cup dk=0$ and $\sum 2\pi n'\cup k\in2\pi\mathbb{Z}$, it follows that $\hat{\mathcal N}\times \hat{\eta}=\hat{\mathcal{N}}$.
By an entirely analogous argument, one finds that
$
\hat{\eta}'\times\hat{\mathcal N}
=
\hat{\mathcal N}.
$

We now compute the fusion rule of $\hat{\mathcal N}$ with its Hermitian
conjugate $\hat{\mathcal{N}}^{\dagger}$.
Acting on the basis state, we obtain
\begin{align}
\hat{\mathcal{N}}^{\dagger}
\hat{\mathcal{N}}
\ket{\{A^{e}\},\{n\}}
&=
\frac{1}{R_{N}^{2}}
\int DA^{e\prime} DA^{e\prime\prime}
\sum_{\{n'\}}\sum_{\{n''\}}
\exp\left[
\frac{iN}{2\pi}\sum
A^{e\prime}\cup
\left(
dA^{e}+2\pi n-dA^{e\prime\prime}-2\pi n''
\right)
\right]
\nonumber\\
&\hspace{3cm}\times
\exp\left[
iN\sum n'\cup (A^{e}-A^{e\prime\prime})
\right]
\ket{\{A^{e\prime\prime}\},\{n''\}}.
\end{align}
The sum over the integer-valued field $n'$ gives
\begin{align}
\sum_{\{n'\}}
\exp\left[
iN\sum n'\cup(A^{e}-A^{e\prime\prime})
\right]
&=
\prod_{(x,\mu)}
\sum_{k_{x,\mu}\in\mathbb Z}
2\pi\,
\delta\!\left(
N(A^{e}_{x,\mu}-A^{e\prime\prime}_{x,\mu})+2\pi k_{x,\mu}
\right)
\nonumber\\
&=
\prod_{(x,\mu)}
\sum_{k_{x,\mu}\in\mathbb Z}
\frac{2\pi}{N}\,
\delta\!\left(
A^{e}_{x,\mu}-A^{e\prime\prime}_{x,\mu}
+\frac{2\pi}{N}k_{x,\mu}
\right).
\end{align}
We decompose the $\mathbb{Z}$-valued 1-form $k$ as
$
k=N\ell+j
$,
where $\ell$ and $k$ are integer valued.  Since $A^{e\prime\prime}$ is integrated over the
fundamental domain $[-\pi,\pi)$, $\ell$ is uniquely fixed,
while $j$ labels the remaining $N$ possible shifts on each link through
\begin{align}
A^{e\prime\prime}=A^{e}+2\pi \ell+\frac{2\pi}{N}j.
\end{align}
Performing the integral for $A^{e\prime\prime}$, we find
\begin{align}
\hat{\mathcal{N}}^{\dagger}
\hat{\mathcal{N}}
\ket{\{A^{e}\},\{n\}}
&=
\frac{1}{R_{N}^{2}}
\left(\frac{2\pi}{N}\right)^{3L^{3}}
\int DA^{e\prime}
\sum_{\{n''\}}
\sum_{\{j\}}
\nonumber\\
&\hspace{1.5cm}\times
\exp\left[
i\sum A^{e\prime}\cup
\left\{
N(n-n''-d\ell)-dj
\right\}
\right]
\ket{
\{A^{e}+2\pi\ell+\frac{2\pi}{N}j\},\,
\{n''\}
},\nonumber\\
&=\frac{1}{R_{N}^{2}}
\left(\frac{2\pi}{N}\right)^{3L^{3}}
\sum_{\{n''\}}
\sum_{\{j\}}2\pi\delta_{N(n-n''-d\ell)-dj,0}
\ket{
\{A^{e}+2\pi\ell+\frac{2\pi}{N}j\},\,
\{n''\}
},
\end{align}
where $\sum_{\{j\}}$ denotes the sum over the $N$ possible values of $j$ on
each link.
The Kronecker delta imposes
\begin{align}
dj=0 \mod N,
\qquad
n''=n-d\ell-\frac{dj}{N} .
\end{align}
Therefore,
\begin{align}
\hat{\mathcal{N}}^{\dagger}
\hat{\mathcal{N}}
\ket{\{A^{e}\},\{n\}}
&=
\frac{(2\pi)^{6L^{3}}}{N^{3L^{3}}R_{N}^{2}}
\sum_{\substack{\{j\}\\
dj=0\ \mathrm{mod}\ N
}}
\ket{
\{A^{e}+2\pi\ell+\frac{2\pi}{N}j\},\,
\{n-d\ell-\frac{dj}{N}\}
}.
\end{align}
The shift by $\ell$ is precisely a $\mathbb{Z}$ 1-form gauge transformation.
Thus, on the physical Hilbert space, where states are invariant under the
$\mathbb Z$ 1-form gauge transformation, this becomes
\begin{align}
\hat{\mathcal{N}}^{\dagger}
\hat{\mathcal{N}}
\ket{\mathrm{phys}}
&=
\frac{(2\pi)^{6L^{3}}}{N^{3L^{3}}R_{N}^{2}}
\sum_{\substack{
\{j\}\\
dj=0\ \mathrm{mod}\ N
}}
\hat{\eta}[j]\ket{\mathrm{phys}}.
\end{align}
We now construct the operator $\hat{\eta}_{\mathbf r}$,
for $\mathbf r=(r_{1},r_{2},r_{3})\in(\mathbb Z_{N})^{3}$,
by specifying the corresponding closed integer-valued $1$-form $k$.
We choose the closed $\mathbb{Z}$-valued $1$-form
\begin{align}
k^{(\mathbf r)}_{x,\mu}
=
r_{1}\delta_{\mu,1}\delta_{x_{1},0}
+
r_{2}\delta_{\mu,2}\delta_{x_{2},0}
+
r_{3}\delta_{\mu,3}\delta_{x_{3},0},
\end{align}
which satisfies
$
d k^{(\mathbf r)}=0.
$
Substituting this into the definition of $\hat{\eta}$, we obtain
\begin{align}
\hat{\eta}_{\mathbf r}
=
\exp\left[
\frac{2\pi i}{N}
\sum_{(x,\mu)}
k^{(\mathbf r)}_{x,\mu}
\left(
\hat{\Pi}^{e}_{x,\mu}
+
\frac{1}{2\pi}(d\hat{A}^{m})_{*(x,\mu)}
\right)
\right].
\end{align}
Choosing the normalization factor
\begin{align}
R_{N}
=
\frac{(2\pi)^{3L^{3}}}{N^{L^{3}}},
\end{align}
and using the fact that the $\mathbb{Z}$ $1$-form gauge transformation acts
trivially on the physical Hilbert space, we obtain the
following fusion rule on physical states
\begin{align}
\hat{\mathcal N}^{\dagger}\times\hat{\mathcal N}
=
\frac{1}{N}
\sum_{r_{1},r_{2},r_{3}=0}^{N-1}
\hat{\eta}_{(r_{1},r_{2},r_{3})}.
\end{align}
which is the expected Tambara--Yamagami type fusion rule. 
Here, we used the fact that
\begin{align}
\sum_{\substack{
\{j\}\\
dj=0\ \mathrm{mod}\ N
}}\hat{\eta}[j]=N^{L^{3}-1}\sum_{r_{1},r_{2},r_{3}=0}^{N-1}\hat{\eta}_{(r_{1},r_{2},r_{3})}
\end{align}
holds on the physical Hilbert space.
This follows because the number of independent degrees of freedom of $j$ satisfying $dj=0\ \mathrm{mod}\ N$ is $L^{3}+2$, while the harmonic part has three independent degrees of freedom. Thus, for each fixed harmonic sector, there are $N^{L^{3}-1}$ exact configurations.

Finally, we discuss how the Hamiltonian is transformed under the defect
$\hat{\mathcal N}$.  In this discussion, we set theta to zero. 
Using the transformation laws \eqref{eq:N transformation for Pi} and \eqref{eq:N transformation for Pi2}, the Hamiltonian \eqref{eq:Maxwell Hamiltonian without theta} transforms as
\begin{align}
H'\hat{\mathcal{N}}=\hat{\mathcal{N}}H,
\end{align}
\begin{align}
\hat H'
=
\frac{1}{2\beta'}
\sum_{(x,\mu)}
\left(\hat{\Pi}^{e\prime}_{x,\mu}\right)^{2}
+
\frac{\beta'}{2}
\sum_{(x,\mu\nu)}
\left(
d\hat A^{e\prime}+2\pi \hat n'
\right)^{2}_{x,\mu\nu},
\end{align}
where $\beta'={N^{2}}/({4\pi^{2}\beta})$.
Thus, the defect $\hat{\mathcal N}$ maps the lattice Maxwell Hamiltonian at
coupling $\beta$ to the Hamiltonian of the same form at coupling
$\beta'=N^{2}/(4\pi^{2}\beta)$. This is the expected transformation of the coupling constant discussed in \cite{Choi:2021kmx,Kan:2024fuu}.

\subsection{Non-invertible defect with the theta term and charges}
Next, we consider the non-invertible defect obtained by gauging the global $\mathbb{Z}_{N}$ $1$-form symmetry in the presence of the theta term and charges.
With the theta term and charges, the non-invertible defect \eqref{eq:non-invertible defect} is modified as
\begin{align}
\hat{\mathcal{N}}^{\rm{T}+\rm{L}}_{\theta}\ket{\{A^{e}\},\{n\}}=\frac{1}{R_{N,\theta}}\int DA^{e\prime}\sum_{\{n'\}}(\mathcal{K}^{\rm{T}}_{\theta}
[A^{e\prime},n';A^{e},n]\mathcal{K}^{\rm{L}}
[A^{e\prime},n';A^{e},n])^{N}\ket{\{A^{e}\},\{n\}},
\end{align}
where $\mathcal{K}^{\rm{T}}_{\theta}$ and $\mathcal{K}^{\rm{L}}$ are defined in \eqref{eq:kernel with theta transverse} and \eqref{eq:kernel longitudinal}.
Using the properties of the projection operators, the transformation of operators under $\hat{\mathcal{N}}^{\rm{T}+\rm{L}}_{\theta}$ is given by
\begin{gather}
\label{eq:Ntheta translation law}
\hat{\mathcal{N}}^{\rm{T}+\rm{L}}_{\theta}P_{\partial}\hat\Pi^{e}_{x,\mu}=-N\sqrt{\beta\beta'}(d\hat{A}^{e\prime}+P_{d}2\pi \hat{n}')_{x-\hat{\mu},\nu\rho}\hat{\mathcal{N}}^{\rm{T}+\rm{L}}_{\theta},\\
\hat{\mathcal{N}}^{\rm{T}+\rm{L}}_{\theta}P_{0}\hat{\Pi}^{e}_{x,\mu}=-\frac{N}{2\pi}(P_{0}2\pi \hat{n}')_{x-\hat{\mu},\nu\rho}\hat{\mathcal{N}}^{\rm{T}+\rm{L}}_{\theta},\\
\hat{\mathcal{N}}^{\rm{T}+\rm{L}}_{\theta}P_{d}\hat{\Pi}^{e}_{x,\mu}=-\frac{N}{2\pi}(P_{\partial}2\pi \hat{n}')_{x-\hat{\mu},\nu\rho}\hat{\mathcal{N}}^{\rm{T}+\rm{L}}_{\theta},\\
P_{\partial}\hat{\Pi}^{e\prime}_{x,\mu}\hat{\mathcal{N}}^{\rm{T}+\rm{L}}_{\theta}=\hat{\mathcal{N}}^{\rm{T}+\rm{L}}_{\theta}N\sqrt{\beta\beta'}(d\hat{A}^{e}+P_{d}2\pi \hat{n})_{x+\hat{\mu},\nu\rho},
\\
P_{0}\hat{\Pi}^{e\prime}_{x,\mu}\hat{\mathcal{N}}^{\rm{T}+\rm{L}}_{\theta}=\hat{\mathcal{N}}^{\rm{T}+\rm{L}}_{\theta}\frac{N}{2\pi}(P_{0}2\pi \hat{n})_{x+\hat{\mu},\nu\rho},
\\
\label{eq:Ntheta translation law end}
P_{d}\hat{\Pi}^{e\prime}_{x,\mu}\hat{\mathcal{N}}^{\rm{T}+\rm{L}}_{\theta}=\hat{\mathcal{N}}^{\rm{T}+\rm{L}}_{\theta}\frac{N}{2\pi}(P_{\partial}2\pi \hat{n})_{x+\hat{\mu},\nu\rho},
\end{gather}
Using these transformation laws, $\hat{\mathcal{N}}^{\rm{T}+\rm{L}}_{\theta}\times\hat{\eta}$ is evaluated as follows
\begin{align}
\hat{\mathcal{N}}^{\rm{T}+\rm{L}}_{\theta}\times\hat{\eta}[k]
&=\hat{\mathcal{N}}^{\rm{T}+\rm{L}}_{\theta}\exp\Bigr[\frac{2\pi i}{N}\sum k_{x,\mu}(P_{d}+P_{\partial}+P_{0})\hat{\Pi}^{e}_{x,\mu}\Bigl]\nonumber\\
&=\exp\Bigr[\frac{2\pi i}{N}\sum(-N\sqrt{\beta\beta'}(d\hat{A}^{e\prime}+P_{d}2\pi n')-\frac{N}{2\pi}(P_{0}+P_{\partial})2\pi n')\cup k\Bigl]\hat{\mathcal{N}}^{\rm{T}+\rm{L}}_{\theta}\nonumber\\
&=\exp\Bigr[-i\sum2\pi n'\cup k\Bigl]\hat{\mathcal{N}}^{\rm{T}+\rm{L}}_{\theta}=\hat{\mathcal{N}}^{\rm{T}+\rm{L}}_{\theta}
\end{align}
In the third line, we used $\sum dA^{e\prime}\cup k=\sum A^{e\prime} \cup dk = 0$ and $\sum P_{d}2\pi n' \cup k = 0$. An entirely analogous calculation shows that $\hat{\eta}^{\prime}\times\hat{\mathcal{N}}^{\rm{T}+\rm{L}}_{\theta}=\hat{\mathcal{N}}^{\rm{T}+\rm{L}}_{\theta}$.
Therefore, even in the presence of the theta term and charges, the fusion rule \eqref{eq:fusion rule eta N} is preserved.

Next, we consider the fusion rule of $\hat{\mathcal{N}}^{\rm{T}+\rm{L}\dagger}_{\theta}\times\hat{\mathcal{N}}^{\rm{T}+\rm{L}}_{\theta}$. By performing a calculation analogous to those in \eqref{eq:S times S appendix1}, \eqref{eq:S times S appendix2}, and \eqref{eq:S times S appendix3} of Appendix~\ref{appendix:Unitarity of S}, we obtain
\begin{align}
&\hat{\mathcal{N}}^{\rm{T}+\rm{L}\dagger}_{\theta}\hat{\mathcal{N}}^{\rm{T}+\rm{L}}_{\theta}\ket{\{A^{e}\},\{n\}}\nonumber\\
&=\frac{1}{R_{N,\theta}^{2}}\int DA^{e\prime\prime}\sum_{\{n''\}}
\prod_{(x,\mu\nu)}2\pi \delta_{N(n-n'')-dk,0}\prod_{(x,\mu)}\sum_{k_{x,\mu}\in\mathbb{Z}}2\pi\delta\Bigr(L_{N}(A^{e}-A^{e\prime\prime}+\frac{2\pi k}{N})\Bigl)\ket{\{A^{e\prime\prime}\},\{n''\}},
\end{align}
where $L_{N}$ is an invertible linear operator defined as
\begin{align}
L_{N}=N(2\pi \sqrt{\beta\beta'}P_{\partial}+P_{0}+P_{d}).
\end{align}
Using $|\det L_{N}|=N^{3L^{3}}(2\pi\sqrt{\beta\beta'})^{2L^{3}-2}$, we obtain
\begin{align}
\hat{\mathcal{N}}^{\rm{T}+\rm{L}\dagger}_{\theta}\hat{\mathcal{N}}^{\rm{T}+\rm{L}}_{\theta}
\ket{\{A^{e}\},\{n\}}
&=
\frac{(2\pi)^{6L^{3}}}{N^{3L^{3}}(2\pi \sqrt{\beta\beta'})^{2L^{3}-2}R_{N,\theta}^{2}}
\sum_{\substack{\{j\}\\
dj=0\ \mathrm{mod}\ N
}}
\ket{
\{A^{e}+2\pi\ell+\frac{2\pi}{N}j\},\,
\{n-d\ell-\frac{dj}{N}\}
}.
\end{align}
Choosing the normalization
\begin{align}
R_{N,\theta}
=
\frac{(2\pi)^{3L^{3}}}{N^{L^3}(2\pi\sqrt{\beta\beta'})^{L^{3}-1}},
\end{align}
and using the fact that the $\mathbb{Z}$ 1-form gauge transformation acts trivially on the physical Hilbert space, we find
\begin{align}
\hat{\mathcal{N}}^{\rm{T}+\rm{L}\dagger}_{\theta}\times\hat{\mathcal{N}}^{\rm{T}+\rm{L}}_{\theta}=\frac{1}{N}
\sum_{r_{1},r_{2},r_{3}=0}^{N-1}
\hat{\eta}_{(r_{1},r_{2},r_{3})}.
\end{align}
on the physical Hilbert space. Therefore, we have shown that the Tambara--Yamagami fusion rule is preserved even in the presence of the theta term and charges.
Finally, let us examine the transformation property of the Hamiltonian \eqref{eq:dynical sector Hamiltonian} and \eqref{eq:charge sector Hamiltonian}. Using Eqs.~\eqref{eq:Ntheta translation law}--\eqref{eq:Ntheta translation law end}, we find
\begin{align}
\hat{\mathcal{N}}^{\rm{T}+\rm{L}}_{\theta}H^{\rm{dyn}}(\beta,\theta)=H^{\rm{dyn}\prime}(\beta',\theta')\hat{\mathcal{N}}^{\rm{T}+\rm{L}}_{\theta}\\
\hat{\mathcal{N}}^{\rm{T}+\rm{L}}_{\theta}H^{\rm{ch}}(\beta,\theta)=H^{\rm{ch}\prime}(\beta',\theta')\hat{\mathcal{N}}^{\rm{T}+\rm{L}}_{\theta}
\end{align}
where 
\begin{gather}
H^{\rm{dyn}\prime}(\beta',\theta')=\frac{1}{2N^{2}\beta'}
\sum
\left(
\hat\Pi_{\mathrm T,x,\mu}^{e\prime}
+
\frac{N^{2}\theta'}{2\pi}\hat h'_{\nu\rho}
\right)^2
+
\frac{N^{2}\beta'}{2}
\sum
\left(
d\hat (A_{\mathrm T}^{e\prime}+2\pi\hat{p})_{x,\mu\nu}
+
2\pi\hat h'_{\mu\nu}
\right)^2,\\
H^{\rm{ch}}(\beta',\theta')=\frac{1}{2N^{2}\beta'}
\sum_{(x,\mu)}
\left(
\hat{\Pi}^{e\prime}_{\mathrm L,x,\mu}
+
\frac{N^{2}\theta'}{2\pi}
\hat n^{m\prime}_{x-\hat\nu-\hat\rho,\nu\rho}
\right)^2
+
\frac{N^{2}\beta'}{2}(2\pi)^2
\sum_{(x,\mu\nu)}
\left(
\hat n^{m\prime}_{x,\mu\nu}
\right)^2.
\end{gather}
Here the transformed parameters $\beta'$ and $\theta'$ are given in \eqref{eq:beta theta transformed}.
From these relations, we find that this non-invertible defect transforms the coupling as
\begin{align}
\mathcal{N}:\tau\longrightarrow-\frac{N^{2}}{\tau}.
\end{align}
This transformation property of the complex coupling constant is consistent with the result of \cite{Choi:2021kmx,Kan:2024fuu}.

\section{Conclusion and discussion}
In this paper, we studied the theta subgroup structure of the lattice
Maxwell theory in the Hamiltonian Villain formulation. We constructed the
$\mathcal{S}$ and $\mathcal{T}^{2}$ transformations at the operator level and showed
that they reproduce the expected modular action on the coupling constant. 
Although a
direct $\mathcal{S}$ transformation does not preserve the original form of the
Hamiltonian with the theta term, the expected modular transformation is recovered after the
topological sector is isolated appropriately. This shows that the Hamiltonian
Villain formulation realizes the expected theta subgroup structure
of Maxwell theory.

We further extended the analysis to sectors with electric and magnetic charges,
introduced as violations of the Gauss-law constraint and the Bianchi identity.
We showed that the $\mathcal{S}$ transformation exchanges electric and magnetic
charges with the expected sign, up to the shift associated with the framing
effect, while the $\mathcal{T}^{2}$ transformation shifts the electric charge by
twice the magnetic charge, realizing the Witten effect in the lattice Hamiltonian
formalism. After accounting for the framing effect, the 
$\mathcal{S}$ transformation gives the expected charge transformation laws.

Finally, we discussed a related non-invertible defect obtained by gauging a
$\mathbb{Z}_{N}$ subgroup of the global $\mathrm{U}(1)$ $1$-form symmetry. We showed that its algebra reproduces the expected Tambara--Yamagami type fusion rule even in the presence of the theta term and electric and magnetic charges.

An important direction for future work is to construct the
$\mathcal{T}$ transformation, rather than the
$\mathcal{T}^{2}$ transformation. A naive $\mathcal{T}$ transformation obtained
by exponentiating the level one lattice Chern--Simons action is not invariant
under the $\mathbb{Z}$ $1$-form gauge symmetry, reflecting the fact that
Chern--Simons theory at odd level is a spin theory. Therefore, to construct the
$\mathcal{T}$ transformation in the Hamiltonian formalism, it would
be necessary to incorporate fermionic degrees of freedom, as in \cite{Xu:2024hyo,Onoda:2025gqa}. It would then be interesting to construct the corresponding
$\mathcal{T}$ transformation, together with an $\mathcal{S}$ transformation that
properly takes into account the fermionic degrees of freedom. This would lead
to a complete understanding of the full
$\mathrm{SL}(2,\mathbb{Z})$ structure of the lattice Maxwell theory in the Villain Hamiltonian formulation. If the $\mathcal{S}$ and $\mathcal{T}$ transformations can be constructed, it may also be possible to construct a triality defect  by $\mathcal{ST}$ gauging \cite{Choi:2022zal}.

Another interesting direction for future work is to clarify the relation between the $\mathcal{T}$ transformation and the discrete chiral transformation in $3+1$d lattice QED. In \cite{Dempsey:2022nys}, it was pointed out that, for the lattice Schwinger model, the continuum fermion mass $m$ and the lattice mass $m_{\mathrm{lat}}$ differ by a finite lattice spacing correction. Taking this mass shift into account, the discrete chiral transformation is realized as a one-site translation of both the staggered fermion and the link variables, under which the theta angle shifts by $\pi$.
To extend this mass shift mechanism to $(3+1)$ dimensions, one may consider $3+1$d lattice QED with staggered fermions. Since staggered fermions in $3+1$d describe two-flavor fermions, the discrete chiral transformation shifts the theta angle by $2\pi$.
In this sense, the discrete chiral transformation in $3+1$d and the $\mathcal{T}$ transformation induce the same transformation of the theta angle.
For free staggered fermions, or when the gauge field is treated as a background field, the operator implementing the discrete chiral transformation can be constructed in terms of one-site translations of the fermions \cite{Aoki:2025vtp}.  Although a construction of the discrete chiral transformation in the presence of a dynamical gauge field is not known, the $\mathcal{T}$ transformation may play an essential role toward such a construction.

\acknowledgments
We would like to thank Y. Ikeda and Y. Furukawa for valuable discussions.
S.A. is supported by RIKEN Special Postdoctoral Researchers Program and JSPS KAKENHI Grant Number 25K17382. T.T. is supported by JST SPRING, Grant Number JPMJSP2108.
Y.K. is supported in part by JSPS KAKENHI Grant Numbers 26K07081 and 26K21723.

\appendix
\section{Differential form on lattice}
\label{app:Differential form}
In this appendix, following ~\cite{Anosova:2022cjm,Aoki:2026pvq}, we summarize the basic facts about differential forms on the lattice. Here, we only present the definitions and relevant properties, for their proofs and further details of the calculations, see \cite{Aoki:2026pvq}.
We consider a $d$-dimensional hypercubic lattice $\Lambda$ on a torus, and denote a lattice site by
$x=(x_{1},x_{2},\ldots,x_{d})$.
An $r$-cell is an oriented elementary hypercube of dimension $r$, specified by a site $x$ and an ordered set of directions
$\mu_{1}\mu_{2}\cdots\mu_{r}$ with
$\mu_{1}<\mu_{2}<\cdots<\mu_{r}$.
An $r$-form $\alpha\in C^{r}(\Lambda,K)$ assigns a value
$\alpha_{x,\mu_{1}\mu_{2}\cdots\mu_{r}}$ to each $r$-cell, where
$K=\mathbb{Z},\mathbb{R}$,$\mathbb{C}$ or $\mathbb{Z}_{N}$.
We take $\alpha$ to be antisymmetric under permutations of its direction indices.

For lattice differential forms, we consider two important differential
operators: the exterior derivative and the boundary operator.
The exterior derivative is defined by
\begin{gather}
d^{(r)}:C^{r}(\Lambda,K)\longrightarrow C^{r+1}(\Lambda,K),\\
(d^{(r)}\alpha)_{x,\mu_{1}\mu_{2}\cdots\mu_{r+1}}
=
\sum_{j=1}^{r+1}
(-1)^{j+1}
\left[
\alpha_{x+\hat{\mu}_{j},
\mu_{1}\cdots \widehat{\mu}_{j}\cdots \mu_{r+1}}
-
\alpha_{x,
\mu_{1}\cdots \widehat{\mu}_{j}\cdots \mu_{r+1}}
\right],
\end{gather}
where the hat over an index means that the corresponding index is omitted.
The boundary operator is defined by
\begin{gather}
\partial^{(r)}:C^{r}(\Lambda,K)\longrightarrow C^{r-1}(\Lambda,K),\\
(\partial^{(r)}\alpha)_{x,\mu_{1}\mu_{2}\cdots\mu_{r-1}}
=
\sum_{\nu=1}^{d}
\left[
\alpha_{x,\mu_{1}\cdots\mu_{r-1}\nu}
-
\alpha_{x-\hat{\nu},\mu_{1}\cdots\mu_{r-1}\nu}
\right].
\end{gather}
When there is no risk of confusion, we omit the superscript indicating the
degree and simply write the operator as $d$ and $\partial$. These operators satisfy nilpotency 
\begin{align}
d^{2}=0,
\qquad
\partial^{2}=0.
\end{align}
It is convenient to use the momentum representation
\begin{align}
\alpha_{x,\mu_{1}\mu_{2}\cdots\mu_{r}}
=
\sum_{p}
\frac{e^{ipx}}{\sqrt{L^{d}}}
\alpha'_{\mu_{1}\mu_{2}\cdots\mu_{r}}(p),
\end{align}
In momentum space, the exterior derivative and the boundary operator act as
\begin{align}
(d\alpha)'_{\mu_{1}\mu_{2}\cdots\mu_{r+1}}(p)
&=
\sum_{j=1}^{r+1}
(-1)^{j+1}
f_{\mu_{j}}(p)\,
\alpha'_{\mu_{1}\cdots \widehat{\mu}_{j}\cdots\mu_{r+1}}(p),
\\
(\partial\alpha)'_{\mu_{1}\mu_{2}\cdots\mu_{r-1}}(p)
&=
\sum_{\nu=1}^{d}
-f_{\nu}(-p)\,
\alpha'_{\mu_{1}\cdots\mu_{r-1}\nu}(p),
\end{align}
where
$
f_{\mu}(p)=e^{ip_{\mu}}-1
$.
Moreover, the following partial integration formula holds 
\begin{align}
\label{eq:partial integration}
\sum_{x}
\sum_{\mu_{1}<\cdots<\mu_{r}}
\alpha_{x,\mu_{1}\cdots\mu_{r}}
(d\beta)_{x,\mu_{1}\cdots\mu_{r}}
=
(-1)^{r}
\sum_{x}
\sum_{\mu_{1}<\cdots<\mu_{r-1}}
(\partial\alpha)_{x,\mu_{1}\cdots\mu_{r-1}}
\beta_{x,\mu_{1}\cdots\mu_{r-1}},
\end{align}
where $\alpha$ is an $r$-form and $\beta$ is an $(r-1)$-form.

We now review the Hodge decomposition on the lattice.
For a $\mathbb{C}$-valued $r$-form $\alpha\in C^{r}(\Lambda,\mathbb{C})$, one can decompose $\alpha$ as
\begin{align}
\alpha=dp+\partial q+h,
\end{align}
where $p$ is an $(r-1)$-form, $q$ is an $(r+1)$-form, and $h$ is a harmonic $r$-form.
Equivalently,
\begin{align}
C^{r}(\Lambda,\mathbb{C})
=
\operatorname{Im} d^{(r-1)}
\oplus
\operatorname{Im} \partial^{(r+1)}
\oplus
H^{r}(\Lambda,\mathbb{C}),
\end{align}
where
\begin{align}
H^{r}(\Lambda,\mathbb{C})
=
\left\{
\alpha\in C^{r}(\Lambda,\mathbb{C})
\,\middle|\,
d\alpha=0,\ \partial\alpha=0
\right\}.
\end{align}
The forms $p$ and $q$ are chosen to satisfy
\begin{align}
p\in\operatorname{Im}\partial^{(r)},
\qquad
q\in\operatorname{Im}d^{(r)}.
\end{align}
With these restrictions, $p$ and $q$ are uniquely determined.

It is useful to introduce projection operators
$P_{d}^{(r)}$, $P_{\partial}^{(r)}$, and $P_{0}^{(r)}$ onto
$\operatorname{Im}d^{(r-1)}$,
$\operatorname{Im}\partial^{(r+1)}$, and
$H^{r}(\Lambda,\mathbb{C})$, respectively.
They are defined by
\begin{align}
\label{eq:definition projection}
P_{d}^{(r)}\alpha=dp,
\qquad
P_{\partial}^{(r)}\alpha=\partial q,
\qquad
P_{0}^{(r)}\alpha=h.
\end{align}
Here, the superscript on each projection operator indicates the degree of the form on which it acts. In what follows, we omit this superscript whenever there is no risk of confusion.
These projection operators satisfy
\begin{align}
P_{d}^+P_{\partial}+P_{0}=1,
\qquad
P_{i}^{2}=P_{i},
\qquad
P_{i}P_{j}=0
\quad
(i\neq j),
\end{align}
where $i,j=d,\partial,0$.
The exact part $dp=P_{d}\alpha$, coexact part $\partial q=P_{\partial}\alpha$, and the harmonic part $h=P_{0}\alpha$ are explicitly given by
\begin{align}
(dp)_{x,\mu_{1}\mu_{2}\cdots\mu_{r}}
&=
(P_{d}\alpha)_{x,\mu_{1}\mu_{2}\cdots\mu_{r}}
\nonumber\\
&=
\sum_{p\neq 0}
\frac{e^{ipx}}{\sqrt{L^{d}}}
\sum_{j=1}^{r}
\sum_{\nu=1}^{d}
\frac{
(-1)^{r-1}
(-1)^{j+1}
f_{\mu_{j}}(p)f_{\nu}(-p)
}{
|f(p)|^{2}
}
\,
\alpha'_{\mu_{1}\cdots \widehat{\mu}_{j}\cdots\mu_{r}\nu}(p).\\
(\partial q)_{x,\mu_{1}\mu_{2}\cdots\mu_{r}}
&=
(P_{\partial}\alpha)_{x,\mu_{1}\mu_{2}\cdots\mu_{r}}
\nonumber\\
&=
\sum_{p\neq 0}
\frac{e^{ipx}}{\sqrt{L^{d}}}
\sum_{\nu=1}^{d}
\sum_{j=1}^{r+1}
\frac{
(-1)^{r}
(-1)^{j+1}
f_{\mu_{j}}(p)f_{\nu}(-p)
}{
|f(p)|^{2}
}
\,
\alpha'_{\mu_{1}\cdots \widehat{\mu}_{j}\cdots\mu_{r}\nu}(p).\\
h_{x,\mu_{1}\mu_{2}\cdots\mu_{r}}
&=(P_{0}\alpha)_{x,\mu_{1}\mu_{2}\cdots\mu_{r}}=
\frac{1}{\sqrt{L^{d}}}
\alpha'_{\mu_{1}\mu_{2}\cdots\mu_{r}}(0).
\end{align}

\section{Cup product}
\label{app:cup product}
In this appendix, we review the definition and the basic properties of the cup product and the
higher cup product on a hypercubic lattice.
The explicit construction of higher cup products on hypercubic lattices was
developed in \cite{Chen:2021ppt}.  We also follow the lattice gauge theory
conventions used in \cite{Jacobson:2023cmr,Aoki:2026pvq}.
These operations provide lattice counterparts of the wedge product.

We denote by $C^{p}(\Lambda,G)$ the set of $G$-valued $p$-forms on a
$d$-dimensional hypercubic lattice $\Lambda$, where $G$ is an abelian group such
as $\mathbb{Z}$, $\mathbb{R}$, $\mathbb{C},$ or $\mathbb{Z}_{N}$.
For a $p$-form $\alpha$ and a $q$-form $\beta$, the ordinary cup product
is a $(p+q)$-form given by
\begin{align}
(\alpha\cup\beta)_{x,\mu_{1}\cdots\mu_{p+q}}
=
\sum_{\sigma\in S_{p+q}}
\epsilon_{\sigma}\,
\frac{1}{p!q!}\,
\alpha_{x,\mu_{\sigma(1)}\cdots\mu_{\sigma(p)}}
\,
\beta_{x+\sum_{i=1}^{p}\hat{\mu}_{\sigma(i)},
\mu_{\sigma(p+1)}\cdots\mu_{\sigma(p+q)}} .
\end{align}
Here $S_{p+q}$ is the symmetric group of degree $p+q$, and $\epsilon_{\sigma}$
is the signature of the permutation $\sigma$.
For example, for a $p$-form $A^{(p)}$ and a $q$-form $B^{(q)}$ in three dimensions,
\begin{align}
&(A^{(1)}\cup B^{(2)})_{x,123}
=
A^{(1)}_{x,1}B^{(2)}_{x+\hat{1},23}
-
A^{(1)}_{x,2}B^{(2)}_{x+\hat{2},13}
+
A^{(1)}_{x,3}B^{(2)}_{x+\hat{3},12},\\
&(B^{(2)}\cup A^{(1)})_{x,123}
=
B^{(2)}_{x,23}A^{(1)}_{x+\hat{2}+\hat{3},1}
-
B^{(2)}_{x,13}A^{(1)}_{x+\hat{1}+\hat{3},2}
+
B^{(2)}_{x,12}A^{(1)}_{x+\hat{1}+\hat{2},3},\\
&(A^{(0)}\cup B^{(3)})_{x,123}=A^{(0)}_{x}B^{(3)}_{x,123},\\
&(B^{(3)}\cup A^{(0)})_{x,123}=B^{(3)}_{x,123}A^{(0)}_{x+\hat{1}+\hat{2}+\hat{3}}.
\end{align}
For a $p$-form $\alpha$ and a $q$-form $\beta$, the cup product satisfies the Leibniz rule
\begin{align}
\label{eq:cup-leibniz}
d(\alpha\cup\beta)
=
d\alpha\cup\beta
+
(-1)^{p}\alpha\cup d\beta .
\end{align}
Using this formula together with integration by parts \eqref{eq:partial integration}, one can show that a $p$-form $\alpha$ and a $(d-p)$-form $\beta$ satisfy
\begin{equation}
\begin{split}
\label{eq:cup and projection}
\sum P_{d}^{(p)}\alpha \cup\beta=\sum\alpha\cup P_{\partial}^{(d-p)}\beta=\sum P_{d}^{(p)}\alpha\cup P_{\partial}^{(d-p)}\beta,\\
\sum P_{\partial}^{(p)}\alpha \cup\beta=\sum\alpha\cup P_{d}^{(d-p)}\beta=\sum P_{\partial}^{(p)}\alpha\cup P_{d}^{(d-p)}\beta,
\end{split}
\end{equation}
where $P_{d}$, $P_{\partial}$, and $P_{0}$ are projection operators defined in \eqref{eq:definition projection}. Note that the projection operators acting on $\alpha$ and $\beta$ generally have different degrees and should therefore be distinguished. In the main text, however, we suppress their degree labels for notational simplicity.

Unlike the continuum wedge product, the lattice cup product is not graded
commutative at the cochain level.  Instead, the failure of graded commutativity
is controlled by the higher cup product $\cup_{1}$.  For
$\alpha\in C^{p}(\Lambda,G)$ and $\beta\in C^{q}(\Lambda,G)$, the product
$\alpha\cup_{1}\beta$ is a $(p+q-1)$-cochain. 
A higher cup product $\cup_1$ on a hypercubic lattice is generally defined in \cite{Chen:2021ppt}. We translate it into the notation of a lattice gauge theory as
\begin{align}
    (\alpha\cup_1 \beta)_{x, \mu_1 \cdots\mu_{p+q-1} } = (-1)^{q}\sum_{a=1}^{p+q-1} (-1)^{a} P_{\mu_a} ( P_{\mu_a} \iota_{\mu_a} \alpha \cup P_{\mu_a} \iota_{\mu_a} \beta  )_{x, \mu_1\cdots \mathring{\mu}_a \cdots \mu_{p+q-1}} \label{eq: higher cup product}.
\end{align}
Here, $P_{\mu} $ is a parity transformation, which acts as 
\begin{align}
    P_{\mu} x =&P_{\mu}(x_1 , \cdots x_{\mu -1}, x_{\mu} , x_{\mu+1},\cdots ,x_d ) \nonumber \\
    =&(x_1 , \cdots x_{\mu -1},- x_{\mu} , -x_{\mu +1},\cdots ,- x_d ),
\end{align}
and
\begin{align}
    (P_{\mu} \alpha)_{x, \mu_1 \cdots \mu_p}=&  \alpha_{P_\mu x, \mu_1 \cdots \mu_{b-1}(-\mu_b) \cdots  (-\mu_p)} \nonumber \\
   =& (-1)^{ p-b+1 }\alpha_{ P_{\mu} x- \sum_{i=b}^{ p } \hat{\mu}_i  ,\mu_1 \cdots \mu_p   } 
\end{align}
with $\mu_{b-1}<\mu \leq \mu_{b}$. $\iota_\mu $ is an interior product given by
\begin{align}
    (\iota_{\mu} \alpha)_{x, \mu_1 \cdots \mu_p}= \alpha_{x,\mu \mu_1 \cdots \mu_p }.
\end{align}
For a $p$-form $A^{(p)}$ and a $q$-form $B^{(q)}$ in three dimensions, the $\cup_{1}$ product is given by
\begin{align}
(A^{(1)}\cup_{1} B^{(3)})_{x,123}
&=
(A^{(1)}_{x,3}+A^{(1)}_{x+\hat{3},2}+A^{(1)}_{x+{\hat{2}}+\hat{3},1})B^{(3)}_{x,123},\\
(B^{(3)}\cup_{1} A^{(1)})_{x,123}
&=
B^{(3)}_{x,123}(A^{(1)}_{x,1}+A^{(1)}_{x+\hat{1},2}+A^{(1)}_{x+{\hat{1}}+\hat{2},3}),\\
(A^{(2)}\cup_{1}B^{(2)})_{x,123}&=A^{(2)}_{x,23}(B^{(2)}_{x,12}+B^{(2)}_{x+\hat{2},13})+A^{(2)}_{x+\hat{2},13}B^{(2)}_{x,12}\nonumber\\
&\quad-A^{(2)}_{x,13}B^{(2)}_{x+\hat{1},23}-A^{(2)}_{x+\hat{3},12}(B^{(2)}_{x,13}+B^{(2)}_{x+\hat{1},23}).
\end{align}
For a $p$-form $\alpha$ and a $q$-form $\beta$, the following identity holds
\begin{align}
\label{eq:cup1-commutativity}
\alpha\cup\beta
-
(-1)^{pq}\beta\cup\alpha
=
(-1)^{p+q+1}
\left[
d(\alpha\cup_{1}\beta)
-
d\alpha\cup_{1}\beta
-
(-1)^{p}\alpha\cup_{1}d\beta
\right].
\end{align}
Thus, the cup product is not strictly graded commutative at the cochain
level.  Instead, its failure to be graded commutative is controlled by the
cup-$1$ product.

\section{Unitarity of the $\mathcal{S}$ transformation with the theta term and charges}
\label{appendix:Unitarity of S}
We now check the unitarity of the $\hat{\mathcal {S}}^{\rm{T}+\rm{L}}_\theta$ defined in \eqref{eq:S with charge}.
The main obstacle in verifying the unitarity of $\hat{\mathcal S}^{\rm{T}+\rm{L}}_\theta$ is the presence of projection operators in the kernel. 
However, by using \eqref{eq:cup and projection}, the projection operator acting on the integration variable can be transferred to the fields on the other side of the cup product.
Acting with $\hat{\mathcal S}^{\rm{T}+\rm{L}\dagger}_{\theta}\hat{\mathcal S}^{\rm{T}+\rm{L}}_\theta$ on the
basis state and using \eqref{eq:cup and projection}, we obtain
\begin{align}
\label{eq:S times S appendix1}
&\hat{\mathcal S}^{\rm{T}+\rm{L}\dagger}_{\theta}\hat{\mathcal S}^{\rm{T}+\rm{L}}_\theta\ket{\{A^{e}\},\{n\}\nonumber}\\
&=
\frac{1}{R_{\theta}^{2}}\int DA^{e\prime}DA^{e\prime\prime}\sum_{\{n'\}}\sum_{\{n''\}}\mathcal{K}^{\rm{T}\dagger}_{\theta}[A^{e\prime\prime},n'';A^{e\prime},n']\mathcal{K}^{\rm{L}\dagger}[A^{e\prime\prime},n'';A^{e\prime},n']\nonumber\\
&\qquad\times\mathcal{K}^{\rm{T}}_{\theta}[A^{e\prime},n';A^{e},n]\mathcal{K}^{\rm{L}}[A^{e\prime},n';A^{e},n]\ket{\{A^{e\prime\prime}\},\{n''\}}\nonumber\\
&=\frac{1}{R_{\theta}^{2}}\int DA^{e\prime}DA^{e\prime\prime}\sum_{\{n'\}}\sum_{\{n''\}}\exp\Bigr[i\sum A^{e\prime}\cup\{\sqrt{\beta\beta'}d(A^{e}-A^{e\prime\prime}+2\pi p-2\pi p'')+(P_{0}+P_{\partial})(n-n'')\}\Bigl]\nonumber\\
&\qquad\times\exp\Bigr[i\sum n'\cup\{2\pi\sqrt{\beta\beta'}P_{\partial}(A^{e}-A^{e\prime\prime}+2\pi p-2\pi p'')+(P_{0}+P_{d})(A^{e}-A^{e\prime\prime})-2\pi P_{\partial}(p-p'')\}\Bigl]\nonumber\\
&\qquad\times\ket{\{A^{e\prime\prime}\},\{n''\}}\nonumber\\
&=\frac{1}{R_{\theta}^{2}}\int DA^{e\prime}DA^{e\prime\prime}\sum_{\{n''\}}\exp\Bigr[i\sum A^{e\prime}\cup\{\sqrt{\beta\beta'}d(A^{e}-A^{e\prime\prime}+2\pi p-2\pi p'')+(P_{0}+P_{\partial})(n-n'')\}\Bigl]\nonumber\\
&\qquad\times\prod_{x,\mu}\sum_{k_{x,\mu}\in\mathbb{Z}}2\pi\delta\Bigr(2\pi\sqrt{\beta\beta'}P_{\partial}(A^{e}-A^{e\prime\prime}+2\pi p-2\pi p'')+(P_{0}+P_{d})(A^{e}-A^{e\prime\prime})-2\pi P_{\partial}(p-p'')+2\pi k\Bigl)\nonumber\\
&\qquad\times\ket{\{A^{e\prime\prime}\},\{n''\}}.\nonumber\\
\end{align}
Since only configurations for which the argument of the delta function vanishes contribute to the integral, we can use this constraint to rewrite the expression as
\begin{align}
\label{eq:S times S appendix2}
&\frac{1}{R_{\theta}^{2}}\int DA^{e\prime}DA^{e\prime\prime}\sum_{\{n''\}}\exp\Bigr[i\sum A^{e\prime}\cup\{d(-\frac{1}{2\pi}(P_{0}+P_{d})(A^{e}-A^{e\prime\prime})+P_{\partial}(p-p'')-k)+(P_{0}+P_{\partial})(n-n'')\}\Bigl]\nonumber\\
&\qquad\times\prod_{x,\mu}\sum_{k_{x,\mu}\in\mathbb{Z}}2\pi\delta\Bigr(2\pi\sqrt{\beta\beta'}P_{\partial}(A^{e}-A^{e\prime\prime}+2\pi p-2\pi p'')+(P_{0}+P_{d})(A^{e}-A^{e\prime\prime})-2\pi P_{\partial}(p-p'')+2\pi k\Bigl)\nonumber\\
&\qquad\times\ket{\{A^{e\prime\prime}\},\{n''\}}\nonumber\\
&=\frac{1}{R_{\theta}^{2}}\int DA^{e\prime}DA^{e\prime\prime}\sum_{\{n''\}}
\exp\Bigr[i\sum A^{e\prime}\cup(n-n''-dk)\Bigl]\nonumber\\
&\qquad\times\prod_{x,\mu}\sum_{k_{x,\mu}\in\mathbb{Z}}2\pi\delta\Bigr(2\pi\sqrt{\beta\beta'}P_{\partial}(A^{e}-A^{e\prime\prime}+2\pi p-2\pi p'')+(P_{0}+P_{d})(A^{e}-A^{e\prime\prime})-2\pi P_{\partial}(p-p'')+2\pi k\Bigl)\nonumber\\
&\qquad\times\ket{\{A^{e\prime\prime}\},\{n''\}}\nonumber\\
&=\frac{1}{R_{\theta}^{2}}\int DA^{e\prime\prime}\sum_{\{n''\}}
\prod_{x,\mu\nu}2\pi \delta_{n-n''-dk,0}\nonumber\\
&\qquad\times\prod_{x,\mu}\sum_{k_{x,\mu}\in\mathbb{Z}}2\pi\delta\Bigr(2\pi\sqrt{\beta\beta'}P_{\partial}(A^{e}-A^{e\prime\prime}+2\pi p-2\pi p'')+(P_{0}+P_{d})(A^{e}-A^{e\prime\prime})-2\pi P_{\partial}(p-p'')+2\pi k\Bigl)\nonumber\\
&\qquad\times\ket{\{A^{e\prime\prime}\},\{n''\}}.
\end{align}
The Kronecker delta imposes $n-n''-dk=0$. Substituting this relation into the Dirac delta function, we obtain 
\begin{align}
\label{eq:S times S appendix3}
&\frac{1}{R_{\theta}^{2}}\int DA^{e\prime\prime}\sum_{\{n''\}}
\prod_{x,\mu\nu}2\pi \delta_{n-n''-dk,0}\nonumber\\
&\qquad\times\prod_{x,\mu}\sum_{k_{x,\mu}\in\mathbb{Z}}2\pi\delta\Bigr(2\pi\sqrt{\beta\beta'}P_{\partial}(A^{e}-A^{e\prime\prime}+2\pi P_{\partial}k)+(P_{0}+P_{d})(A^{e}-A^{e\prime\prime})-2\pi P_{\partial}k+2\pi k\Bigl)\nonumber\\
&\qquad\times\ket{\{A^{e\prime\prime}\},\{n''\}}\nonumber\\
&=\frac{1}{R_{\theta}^{2}}\int DA^{e\prime\prime}\sum_{\{n''\}}
\prod_{x,\mu\nu}2\pi \delta_{n-n''-dk,0}\nonumber\\
&\qquad\times\prod_{x,\mu}\sum_{k_{x,\mu}\in\mathbb{Z}}2\pi\delta\Bigr(2\pi\sqrt{\beta\beta'}P_{\partial}(A^{e}-A^{e\prime\prime}+2\pi k)+(P_{0}+P_{d})(A^{e}-A^{e\prime\prime}+2\pi k)\Bigl)\nonumber\\
&\qquad\times\ket{\{A^{e\prime\prime}\},\{n''\}}\nonumber\\
&=\frac{1}{R_{\theta}^{2}}\int DA^{e\prime\prime}\sum_{\{n''\}}
\prod_{x,\mu\nu}2\pi \delta_{n-n''-dk,0}\prod_{x,\mu}\sum_{k_{x,\mu}\in\mathbb{Z}}2\pi\delta\Bigr(L(A^{e}-A^{e\prime\prime}+2\pi k)\Bigl)\ket{\{A^{e\prime\prime}\},\{n''\}}\nonumber\\
\end{align}
where $L$ is an invertible linear operator defined as
\begin{align}
L=2\pi \sqrt{\beta\beta'}P_{\partial}+P_{0}+P_{d}.
\end{align}
Therefore, we obtain
\begin{align}
\hat{\mathcal S}^{\rm{T}+\rm{L}\dagger}_{\theta}\hat{\mathcal S}^{\rm{T}+\rm{L}}_\theta\ket{\{A^{e}\},\{n\}\nonumber}
&=\frac{(2\pi)^{6L^3}}{R^{2}_{\theta}|\det L|}\ket{\{A^{e}+2\pi k\},\{n-dk\}}\nonumber\\
&=\frac{(2\pi)^{6L^3}}{R^{2}_{\theta}(2\pi\sqrt{\beta\beta'})^{2L^{3}-2}}\ket{\{A^{e}+2\pi k\},\{n-dk\}}.\nonumber\\
\end{align}
In the last line, we used the fact that the number of degrees of freedom in the transverse sector is $2L^{3}-2$. 
Taking $R_{\theta}=\frac{(2\pi)^{3L^{3}}}{(2\pi \sqrt{\beta\beta'})^{L^{3}-1}}$, and using the fact that the $\mathbb{Z}$ $1$-form gauge transformation acts trivially on the physical Hilbert space, we conclude that $\hat{\mathcal {S}}^{\rm{T}+\rm{L}}_{\theta}$ is unitary on the physical Hilbert space.


\bibliographystyle{JHEP}
\bibliography{ref}


\end{document}